\pdfoutput=1
\documentclass[11pt]{article}
\usepackage{jheppub}

\usepackage{amsmath}
\usepackage{amssymb}
\usepackage{graphicx,color}
\usepackage{amsthm}
\usepackage{mathrsfs}

\usepackage{tikz}
\usetikzlibrary{matrix,arrows}

\usepackage{ifpdf}

\usepackage{subfigure}

\usepackage[boxsize=0.5em,aligntableaux=center]{ytableau}

\usepackage{tikz}
\usetikzlibrary{shapes,shadows,calc}
\usepgflibrary{arrows}
\usepackage{rotating}
\usepackage{wrapfig}

\tikzset{
  sshadow/.style={opacity=.25, shadow xshift=0.05, shadow yshift=-0.06},
}

\def\kbbox[#1,#2,#3,#4,#5]#6{
        \draw[dashed] node[draw,color=gray!50,minimum
        height=#1,minimum width=#2] (#4) at #5 {}; 
        \node[anchor=#3,inner sep=2pt] at (#4.#3)  {#6};
}
\def\kbboxred[#1,#2,#3,#4,#5]#6{
        \draw[] node[draw,color=red,minimum
        height=#1,minimum width=#2] (#4) at #5 {}; 
        \node[anchor=#3,inner sep=2pt] at (#4.#3)  {#6};
}

\newcommand{\rep}[1]{\mathbf{#1}}

\newcommand{\tk}{\tilde{k}}
\newcommand{\tq}{\tilde{q}}

\newcommand{\bC}{\mathbb{C}}

\newcommand{\bP}{\mathbb{P}}

\newcommand{\bZ}{\mathbb{Z}}

\newcommand{\cC}{\mathcal{C}}

\newcommand{\cK}{\mathcal{K}}
\newcommand{\cL}{\mathcal{L}}

\newcommand{\cN}{\mathcal{N}}
\newcommand{\cO}{\mathcal{O}}

\newcommand{\cR}{\mathcal{R}}

\newcommand{\cT}{\mathcal{T}}
\newcommand{\cV}{\mathcal{V}}

\newcommand{\sG}{\mathsf{G}}

\newcommand{\sS}{\mathsf{S}}

\newcommand{\wt}[1]{\widetilde{#1}}
\newcommand{\tA}{\widetilde{A}}

\newcommand{\tX}{\mathcal{X}}

\newcommand{\wh}[1]{\widehat{#1}}
\newcommand{\hX}{\mathbb{X}}

\DeclareMathOperator{\Tr}{Tr}

\DeclareMathOperator{\sign}{sign}
\DeclareMathOperator{\tr}{tr}
\DeclareMathOperator{\Res}{Res}

\def\Label#1{\label{#1}%
  \smash{\hbox to0pt{\raise1ex\hbox{\tiny[#1]}\hss}}}
\def\noLabels{\let\Label=\label}
\def\nobbibitem{\let\bbibitem=\bibitem}
 \def\noBibitem{\let\Bibitem=\bibitem}

\newcommand{\be}{\begin{equation}}
\newcommand{\ee}{\end{equation}}
\newcommand{\beq}{\begin{equation}}
\newcommand{\eeq}{\end{equation}}
\newcommand{\bea}{\begin{eqnarray}}
\newcommand{\eea}{\end{eqnarray}}
\newcommand{\R}{\text{Re}}
\newcommand{\I}{\text{Im}}

\newcommand{\eref}[1]{(\ref{#1})}

\newcommand{\ba}{\begin{eqnarray}}
\newcommand{\ea}{\end{eqnarray}}
\newcommand{\nn}{\nonumber}

\newcommand\varpm{\mathbin{\vcenter{\hbox{%
  \oalign{\hfil$\scriptstyle+$\hfil\cr
          \noalign{\kern-.3ex}
          $\scriptscriptstyle({-})$\cr}%
}}}}
\newcommand\varmp{\mathbin{\vcenter{\hbox{%
  \oalign{$\scriptstyle({+})$\cr
          \noalign{\kern-.3ex}
          \hfil$\scriptscriptstyle-$\hfil\cr}%
}}}}

\title{\centering Physics of F-theory compactifications without section}

\author[a]{Lara B. Anderson,}
\author[b]{I\~naki Garc\'ia-Etxebarria,}
\author[b]{Thomas W. Grimm,}
\author[b]{and Jan Keitel}

\affiliation[a]{Department of Physics, Robeson Hall, 0435, Virginia Tech,\\
850 West Campus Drive, Blacksburg, VA 24061, USA}
\affiliation[b]{Max Planck Institute for Physics,\\
F\"ohringer Ring 6, 80805 Munich, Germany}

\emailAdd{lara.anderson@vt.edu}
\emailAdd{inaki@mpp.mpg.de}
\emailAdd{grimm@mpp.mpg.de}
\emailAdd{jkeitel@mpp.mpg.de}

\abstract{We study the physics of F-theory compactifications on
  genus-one fibrations without section by using an M-theory dual
  description. The five-dimensional action obtained by considering
  M-theory on a Calabi-Yau threefold is compared with a
  six-dimensional F-theory effective action reduced on an additional
  circle. We propose that the six-dimensional effective action of
  these setups admits geometrically massive $U(1)$ vectors with a
  charged hypermultiplet spectrum.  The absence of a section induces
  NS-NS and R-R three-form fluxes in F-theory that are non-trivially
  supported along the circle and induce a shift-gauging of certain
  axions with respect to the Kaluza-Klein vector.  In the
  five-dimensional effective theory the Kaluza-Klein vector and the
  massive $U(1)$s combine into a linear combination that is
  massless. This $U(1)$ is identified with the massless $U(1)$
  corresponding to the multi-section of the Calabi-Yau threefold in
  M-theory.  We confirm this interpretation by computing the one-loop
  Chern-Simons terms for the massless vectors of the five-dimensional
  setup by integrating out all massive states. A closed formula is
  found that accounts for the hypermultiplets charged under the
  massive $U(1)$s.}

\begin{document}
\setlength{\parskip}{5pt}

\makeatletter
\let\old@fpheader\@fpheader
\renewcommand{\@fpheader}{\old@fpheader\hfill
MPP-2014-258}
\makeatother

\maketitle
\newpage

\section{Introduction}

F-theory, as introduced in \cite{Vafa:1996xn}, provides a beautiful 
geometric reformulation of Type IIB string theory with varying string 
coupling. Not only has it been explored from a formal perspective, but, 
more recently, it has also found exciting applications to realistic
model building, starting with
\cite{Donagi:2008ca,Beasley:2008dc,Hayashi:2008ba,Beasley:2008kw}. 
The underlying idea of F-theory is to identify the complexified 
string coupling $\tau$ of Type IIB string theory with the complex structure 
of an auxiliary two-torus. Such an interpretation is motivated by the existence of 
the non-perturbative $SL(2,\mathbb{Z})$ symmetry of Type IIB. 
Remarkably, this construction extends to situations in which $\tau$
depends non-trivially on the space-time coordinates of the Type IIB background. 
One can thus consider backgrounds in which the $T^2$ is fibered over 
some compact base manifold. If the effective theory is to be supersymmetric 
the entire $T^2$ fibration $X$ must be a Calabi-Yau manifold.

So far, most of the literature has focused on a subclass of $T^2$ fibrations $X$
that are simpler to analyze. Namely, it has largely been assumed that $X$
has a section, that is, a global meromorphic embedding of the base
into the total space of the fibration; or equivalently, a canonical
choice of point in the fiber well defined everywhere (except possibly
at some lower-dimensional loci in the base where the fiber degenerates). All such
fibrations can be birationally transformed~\cite{Nakayama} into a
\emph{Weierstrass model} of the form
\begin{align}
  \label{eq:Weierstrass}
  y^2 = x^3 + fxz^4 + gz^6
\end{align}
with $(x,y,z)$ coordinates of a $\bP^{2,3,1}$, and $f,g$ functions on
the base of the fibration. A canonical section is provided by
$z=0$.
As pointed out by Witten in \cite{Witten:1996bn}, this subclass of models
is physically simpler to treat, because the existence of a section implies the
absence of certain fluxes, as we will explain in more detail later on.
Geometrically, the restriction to Weierstrass models facilitated
model building with non-Abelian gauge symmetries, as the widely
used algorithm of \cite{Bershadsky:1996nh} (see also \cite{Katz:2011qp, Lawrie:2012gg}
for later extensions) could be applied
directly to models with Weierstrass form.

We emphasize, however, that while the assumption of having a section
simplifies the analysis, it is in no way necessary for the consistency
of the physics, or the existence of an F-theory limit. In fact, it is
very easy to construct $T^2$ fibrations with no section that serve
as natural backgrounds for F-theory and we analyze explicitly various
examples below. For completeness, let us also note that the approach
taken by \cite{Candelas:1996su, Bouchard:2003bu} provides a convenient
and more general way of generating non-Abelian gauge symmetries
also for models without section.

Based on this observation, in this paper we want to explore the
physics of F-theory backgrounds $\tX$ in which the $T^2$ does \emph{not}
have a section, and thus no Weierstrass model. This case remains
basically unexplored, with the exception of the recent
works~\cite{Braun:2014oya,Morrison:2014era} (which appeared while this
work was in progress), and some remarks in \cite{Witten:1996bn} that
will play a role in our analysis below. We will focus on the formal
aspects of this class of F-theory backgrounds, uncovering some
interesting characteristics of the resulting effective field theories.

We will argue that a \emph{massive} $U(1)$ symmetry in the resulting
six-dimensional theory coming from F-theory on $\tX$ plays an
essential role in a proper understanding of the theory. In fact, one
of the important results in this paper is a proposal for a method of
computing the massless and part of the massive spectrum of F-theory on
a fibration $\tX$ without section. We will test this proposal in a
particular class of examples where the origin and properties of this
massive $U(1)$ are particularly transparent --- namely, examples where
$\tX$ is obtained from a conifold transition from a Calabi-Yau
threefold $\hX$ with \emph{two} sections. Note that massive $U(1)$s
in F-theory have recently been investigated in
\cite{Grimm:2010ez,Grimm:2011tb,Braun:2014nva}.

In fact, for the cases studied in detail in this paper there exist
both geometrical and physical reasons for why the Calabi-Yau manifolds
$\tX$ with bi-section are naturally related to fibrations $\hX$ with
two independent sections. Geometrically, by transitioning to a
different manifold $\hX$ the bi-section can be split into two
independent sections. Physically, the massive $U(1)$ becomes massless
in that limit.  Recently, the study of massless $U(1)$ gauge
symmetries in global F-theory compactifications has been a heavily
investigated topic. Geometrically, the number of the Abelian gauge
fields corresponds to the rank of the Mordell-Weil group of the
fibration. As the Mordell-Weil group is generated by the sections,
there is a direct correspondence between the number of independent
sections and the number of $U(1)$ generators. Let us note here that
starting with the $U(1)$-restricted models of \cite{Grimm:2010ez},
continued by a systematic six-dimensional analysis of single $U(1)$
models \cite{Morrison:2012ei} and extended to more general treatments
of multiple $U(1)$ factors \cite{Mayrhofer:2012zy, Borchmann:2013jwa,
  Cvetic:2013nia, Grimm:2013oga, Braun:2013nqa, Cvetic:2013uta,
  Borchmann:2013hta, Cvetic:2013jta, Cvetic:2013qsa} both with
holomorphic and non-holomorphic sections \cite{Braun:2013yti,
  Cvetic:2013nia, Grimm:2013oga} a variety of methods has been
developed that we will draw from in order to analyze the properties of
our specific models.

However, in order to study the effective physics of the F-theory compactifications
without sections, it is most useful to employ the M- to F-theory limit.
One can define F-theory on a $T^2$ fibered manifold $X$ as M-theory
compactified on $X$ in the limit where the size of the $T^2$ fiber
goes to 0. When the $T^2$ is small, but of finite size, F-theory is
compactified on $X\times S^1$, with the size of the $S^1$ inversely
proportional to the area of the $T^2$ fiber (so in the strict F-theory
limit the $S^1$ decompactifies). Much of the subtle behavior of
F-theory on manifolds $\tX$ without a section can be best understood by taking
the $S^1$ to have finite size. For concreteness, in this paper we take
$\dim_\bC(\tX)=3$, so F-theory on $\tX$ gives a six-dimensional
theory. Further compactification on an $S^1$ gives a five-dimensional
theory, which can be alternatively obtained by compactifying M-theory
on $\tX$. Matching the two five-dimensional theories then allows one to identify
geometric quantities of $\tX$ with physical observables of
the effective F-theory physics \cite{Grimm:2010ks, Bonetti:2011mw}.

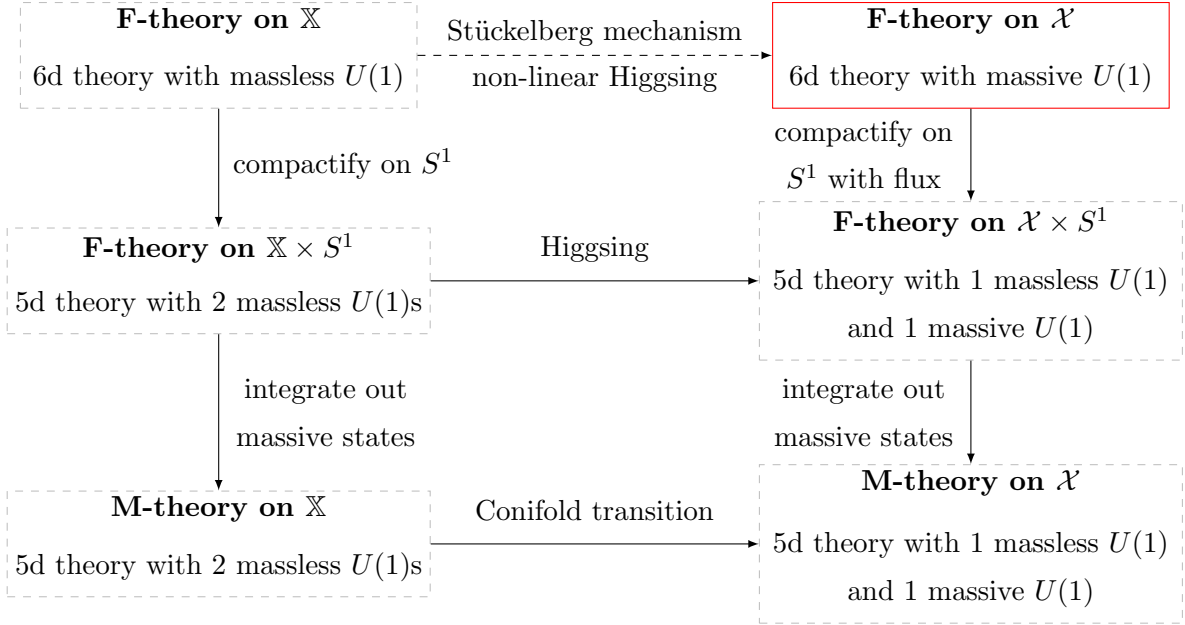
\begin{figure}
\centering
  \begin{tikzpicture}
    
  \kbbox[40, 150,north,F_resolved,(0, 0)] {\textbf{F-theory on $\hX$}};
    \node [] () at (0, -0.3) {6d theory with massless $U(1)$};
  
  \kbboxred[40, 150,north,F_deformed,(10, 0)] {\textbf{F-theory on $\tX$}};
    \node [] () at (10, -0.3) {6d theory with massive $U(1)$};
    
  \draw[dashed, -latex] (F_resolved.east) -> 
  node[above]{St\"uckelberg mechanism} node[below]{non-linear Higgsing}  (F_deformed.west);
  
  \kbbox[40, 160,north,F_resolved_S1,(0, -3)] {\textbf{F-theory on $\hX \times S^1$}};
    \node [] () at (0, -3.3) {5d theory with $2$ massless $U(1)$s};
  
  \kbbox[60, 160,north,F_deformed_S1,(10, -3)] {\textbf{F-theory on $\tX \times S^1$}};
    \node [] () at (10, -3.3) {\begin{tabular}{c}
    5d theory with $1$ massless $U(1)$ \\
    and $1$ massive $U(1)$\end{tabular}};

  \draw [-latex] (F_resolved.south) -> node[above, yshift=-0.7em, xshift=4.3em]{
  compactify on $S^1$}(F_resolved_S1.north);
  
  \draw [-latex] (F_deformed.south) -> node[above, yshift=-2em, xshift=-3.7em]{\begin{turn}{0}
 \begin{tabular}{c}compactify on\\$S^1$ with flux\end{tabular}
 \end{turn}}(F_deformed_S1.north);
  
  \kbbox[40, 160,north,M_resolved,(0, -6.5)] {\textbf{M-theory on $\hX$}};
    \node [] () at (0, -6.8) {5d theory with $2$ massless $U(1)$s};
  
  \kbbox[60, 160,north,M_deformed,(10, -6.5)] {\textbf{M-theory on $\tX$}};
    \node [] () at (10, -6.8) {\begin{tabular}{c}
    5d theory with $1$ massless $U(1)$ \\
    and $1$ massive $U(1)$\end{tabular}};
    
 \draw [-latex] (F_resolved_S1.south) -> node[above, yshift=-2em, xshift=3.7em]{\begin{turn}{0}
 \begin{tabular}{c}integrate out \\ massive states \end{tabular}
 \end{turn}}(M_resolved.north);
 \draw [-latex] (F_deformed_S1.south) -> node[above, yshift=-2em, xshift=-3.7em]{\begin{turn}{0}
 \begin{tabular}{c}integrate out \\ massive states \end{tabular}
 \end{turn}} (M_deformed.north);
 
 \draw [-latex] (F_resolved_S1.east) -> node[above, yshift=0em, xshift=0em]{\begin{turn}{0}
 \begin{tabular}{c}Higgsing\end{tabular}
 \end{turn}}
 (F_deformed_S1.west);
  \draw [-latex] (M_resolved.east) -> node[above, yshift=0em, xshift=0em]{\begin{turn}{0}
 \begin{tabular}{c}Conifold transition\end{tabular}
 \end{turn}}
 (M_deformed.west);
  
  \end{tikzpicture}
  \caption{Overview of our discussion. The object of interest in the
    top-right corner, corresponding to the six-dimensional theories coming from
    F-theory on a space without section $\tX$. In the examples we will
    discuss explicitly these compactifications are closely related (by
    making some fields massive) to F-theory on spaces with section
    $\hX$, giving the six-dimensional theories in the top-left
    corner. Compactification of these theories on $S^1$ gives two five-dimensional 
    theories, in the middle row, which can also be obtained by
    M-theory on the corresponding Calabi-Yau threefolds (shown in 
    the bottom row). The five-dimensional
    theories are related by Higgsing, or equivalently, by conifold
    transitions in M-theory.}
  
  \label{fig:overview}
  \end{figure}

We have organized this paper as follows. Section~\ref{sec:6d} contains
a general discussion of the six-dimensional theories arising from
F-theory compactifications on $T^2$-fibered Calabi-Yau threefolds with
no section. Section~\ref{sec:reduction} then describes the
reduction of these theories down to five dimensions by
compactification on a circle. A number of subtleties arise, which we
solve. This general discussion is then illustrated in
section~\ref{sec:examples} in a number of examples. Since there are a
number of different actors in play in our construction, we have
summarized the outline of our discussion in figure~\ref{fig:overview}
for the convenience of the reader.

\section{Six-dimensional action of F-theory on multi-section threefolds}
\label{sec:6d}
\begin{figure}[h!]
\centering
  \begin{tikzpicture}
    
    \kbbox[90, 145,north,F_resolved,(0, 0)] {\textbf{F-theory on $\hX$}};
    \node [] () at (-1, 0.6) {Massless sector:};
    \node [] () at (0.2, -0.6) {\begin{tabular}{ll}
     1 & gauge field $\hat A^1$\\
     $H_{U(1)}$ & charged hypers \\
     $H_{neutral}$ & neutral hypers \\
  \end{tabular}};
  
  \kbboxred[90, 170,north,F_deformed,(10, 0)] {\textbf{F-theory on $\tX$}};
    \node [] () at (8.6, 0.6) {Massless sector:};
    \node [] () at (10.2, -0.2) {\begin{tabular}{ll}
     $H_{U(1)}-1$ & charged hypers \\
     $H_{neutral}$ & neutral hypers \\
  \end{tabular}};
    \node [] () at (9.4, -1.3) {$1$ massive gauge field $\hat A^1$};
    
  \draw[dashed, -latex] (F_resolved.east) -> 
  node[above]{St\"uckelberg mechanism} node[below]{non-linear Higgsing} (F_deformed.west);  
  \end{tikzpicture}
  \caption{Six-dimensional effective theories with a massless and massive $U(1)$ gauge field.} \label{fig:6dstory}
  \end{figure}
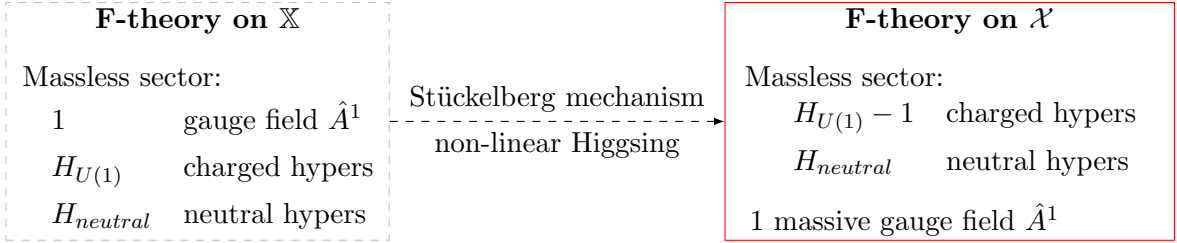

In this section we introduce the six-dimensional effective theories 
that we claim to arise in F-theory compactifications on a genus-one 
fibered Calabi-Yau threefold $\tX$ with a multi-section. 
To begin with, we recall in subsection \ref{sec:6dmasslessU(1)} the effective theory of an F-theory 
compactification on a manifold $\hX$ with two sections. This theory will admit a massless 
Abelian gauge field $\hat A^1$, where the hat indicates here and in the following 
that we are dealing with a field in a six-dimensional space-time. In contrast, we explain in subsection \ref{sec:6dmassiveU(1)}
that the compactification on $\tX$ yields a $U(1)$ gauge 
field $\hat A^1$ made massive by a St\"uckelberg mechanism. 
For simplicity, we will restrict ourselves 
to scenarios with a single Abelian gauge field and no non-Abelian gauge symmetry. 
In geometric terms this amounts to assuming that  
$\tX$ has a bi-section, i.e.~a multi-section of rank two, and no non-Abelian 
singularities. We discuss the first row in figure~\ref{fig:overview} and thus establish figure~\ref{fig:6dstory}.

\subsection{Review of massless U(1) in F-theory} \label{sec:6dmasslessU(1)}

In order to set the stage for our considerations of a massive $U(1)$, let us first 
recall the simpler situation in which the $U(1)$ is massless. Six-dimensional effective 
theories with a massless $U(1)$ arise when considering F-theory on a 
manifold with two sections $\hX$. One of these sections is identified 
with the massless $U(1)$ while the second section, the zero section,
corresponds to the Kaluza-Klein vector in the F-theory to M-theory reduction as 
we recall in section \ref{sec:reduction}. The effective theory for F-theory compactifications 
with multiple sections was studied in detail in \cite{Grimm:2013oga}. 
The spectrum of the six-dimensional theory 
consists of $T$ tensor multiplets and $V$ vector multiplets with 
\begin{equation}
   T= h^{1,1}(B_2) -1 \ , \qquad V = h^{1,1}(\hX) - h^{1,1}(B_2) - 1 = 1\ ,
\end{equation}
where we have considered, for simplicity, that $\hX$ induces no non-Abelian 
gauge symmetries. The base of the elliptic fibration is denoted by $B_2$.
The vector multiplet contains precisely the massless $U(1)$ vector $\hat A^1$.
In addition to these multiplets the theory will generally contain a number of hypermultiplets 
$H$. Generally, one can split 
\begin{equation}
    H = H_{\rm neutral} + H_{\rm charged} \ , \qquad H_{\rm neutral} = h^{2,1} (\hX)+ 1\
\end{equation}
and if there are no non-Abelian gauge symmetries
\begin{align}
 H_{\rm charged} = H_{U(1)}\,,
\end{align}
where $H_{U(1)}$ counts the number of hypermultiplets charged under $\hat A^1$.
Recall that the cancellation of six-dimensional pure gravitational anomalies requires
the relation 
\beq \label{grav_anomaly}
      H - V  = 273 - 29 T\ . 
\eeq
In addition one has to cancel the gauge and mixed anomalies. In order 
to do that one can employ a generalized Green-Schwarz mechanism \cite{Green:1984sg,Sagnotti:1992qw,Sadov:1996zm}
induced by a coupling
\begin{equation} \label{SGS}
   S_{\rm GS} =
    - \frac{1}{2} \int \Omega_{\alpha \beta}  \hat B^\alpha \wedge \Big( \frac{1}{2} a^\beta
    \Tr (\mathcal{\hat R} \wedge \mathcal{\hat R})  + 2 b^\alpha \hat F^1 \wedge \hat F^1\Big)\,,
\end{equation}
where $\mathcal{\hat R}$ is the six-dimensional curvature two-form and
$\hat F^1$ is the field strength of the U(1) vector $\hat A^1$.  The
tensors $\hat B^\alpha$, $\alpha =1,...,T+1$ arise from the $T$ tensor
multiplets and the gravity multiplet, and the symmetric constant
matrix $\Omega_{\alpha \beta}$ and the constant vectors $(a^\alpha,
b^\alpha)$ are crucial to determine the couplings of the
six-dimensional supergravity theory.

Finally, recall that both $(a^\alpha, b^\alpha)$ and $\Omega_{\alpha \beta}$ are
naturally determined by the topology of the compactification manifold $\hX$ as
\begin{align}
 a^\alpha = -\Omega^{\alpha \beta} \Big( D_{\beta} \cdot [\pi^{*} c_1(B_2)] \Big)_{B_2}\,, \quad
 b^\alpha = - \Omega^{\alpha \beta} \left(D_{U(1)}^2 \cdot D_{\beta}\right) \,, \quad 
 \Omega_{\alpha \beta} = \left(D_{\alpha} \cdot D_{\beta} \right)_{B_2}\,,
\end{align}
where we have denoted by $D_{\alpha}$ the divisors inside $\hX$ that are obtained
by fibering the genus-one curve over a divisor in the base $B_2$ and 
write $[\pi^* c_1(B_2)]$ for the Poincar\'e-dual of the first Chern class of $B_2$ pulled back
to $\hX$. Furthermore, we take $\Omega^{\alpha \beta}$
to be the inverse of $\Omega_{\alpha \beta}$. $D_{U(1)}$ is the divisor in $\hX$ obtained 
from the $U(1)$ seven-brane divisor in the base.

\subsection{Massive $U(1)$ and the St\"uckelberg mechanism} \label{sec:6dmassiveU(1)}

Let us now turn to the compactifications most relevant to this work
and consider F-theory on the space $\tX$ with bi-section.  We propose
that in this case one finds a massive $U(1)$ vector multiplet that can
be described by a massless $U(1)$ vector multiplet coupled to a
hypermultiplet by a St\"uckelberg mechanism.  In addition to this
non-linearly charged hypermultiplet, $H_{U(1)}-1$ matter
hypermultiplets will be part of the six-dimensional effective theory.
In the next section we will use the dual M-theory picture in order to
argue for the correctness of this proposal.  We will also determine
the total number of charged hypermultiplets $H_{U(1)}$ and their
charges.

Let us denote the scalars in the $H_{U(1)}-1$ linearly charged matter hypermultiplets by 
$h^s$. The additional non-linearly charged hypermultiplet contains an axion 
$c$ with shift symmetry gauged under $\hat A^1$. In summary, 
one has\footnote{Since the scalars $c,h^s$ remain scalars without redefinition when compactifying 
the theory to five dimensions in section \ref{sec:reduction}, we will slightly abuse notation and not put a hat on them to 
distinguish them from their five-dimensional counterparts.}
\beq \label{gauged_U1}
   \mathcal{\hat D} c =  d c + m \hat A^1\ , \qquad \quad \mathcal{\hat D} h^s= d h^s + q^s \, \hat A^1 h^s\ , 
\eeq
where $q^s$ is the charge of the state $h^s$. In other words, the 
theory differs from the one introduced in the previous subsection \ref{sec:6dmasslessU(1)} due
to the gauging of the shift symmetry of $c$ parametrized by $m$. More details on the difference 
between the non-linear Higgs mechanism induced by the coupling to 
$c$ and a linear Higgs mechanism are discussed in \cite{Garcia-Etxebarria:2014qua}.

After gauge fixing the $U(1)$ gauge symmetry, the kinetic term $|\mathcal{\hat D}c|^2$ of the axion $c$ becomes
a mass term for $\hat A^1$, which is proportional to $m^2$. Hence, the $U(1)$
can become massive by ``eating'' the axion $c$. 
In F-theory the shift gauging \eqref{gauged_U1} can arise from a geometric 
St\"uckelberg mechanism \cite{Grimm:2011tb}. More precisely, if the seven-brane 
action induces a six-dimensional coupling 
\beq \label{Stueck}
  S_{\rm St} =   \int_{M^{5,1}} m\,  c_4 \wedge \hat F^1 \ , 
\eeq
then the four-form $c_4$ can be dualized into the axion $c$ to obtain the gauging \eqref{gauged_U1}.

For D7-branes at weak coupling the effective coupling \eqref{Stueck} arises indeed from 
a non-trivial Chern-Simons coupling 
$ \int_{\mathcal{M}_8} C_6 \wedge F$,
where $C_6$ is the R-R six-form of Type IIB string theory, and 
$\mathcal{M}_8 = M^{5,1} \times \mathcal{C}^{\rm D7}$ is the eight-dimensional subspace 
wrapped by the D7-brane and its orientifold image \cite{Jockers:2004yj}. 
Comparing \eqref{Stueck} with these Chern-Simons terms one finds
$m\, c_4= \int_{\mathcal{C}^{\rm D7}} C_6$, which determines $m$ as 
an intersection number at weak string coupling. Since the axion 
$c$ is the dual of $c_4$ in six dimensions, it arises in the expansion of the 
R-R two-from $C_2$ as
\beq \label{C2expansion}
    C_2 = c \, \tilde \omega\ ,
\eeq
where $\tilde \omega$ is a $(1,1)$-form on the Type IIB covering space that 
is negative under the orientifold involution. 
Since there is no flux 
involved in this mechanism, it was termed geometric St\"uckelberg mechanism
in \cite{Grimm:2011tb}. It should be stressed that determining $m$ in a
general F-theory setting is more involved and we will return to 
this question in the later parts of the paper. 

For completeness, let us consider the effective theory at an energy scale below the 
mass of the $U(1)$. In order to obtain this theory we have to integrate out 
the massive vector multiplet containing $A_1$, which was obtained 
by a massless vector multiplet ``eating'' a massless hypermultiplet. In other 
words one finds 
\beq
   V \ \rightarrow \ V - 1\ , \qquad H \ \rightarrow \ H -1\ ,  
\eeq
consistent with \eqref{grav_anomaly}. Furthermore, all hypermultiplets 
charged under the massive $U(1)$ are neutral in the effective theory
and one has 
\beq
   H_{\rm charged}  \ \rightarrow \  0 \ , \qquad H_{\rm neutral}     \rightarrow \  H_{\rm neutral} + H_{U(1)}-1\ .
\eeq
While this theory is a valid effective theory at the massless level, we will see in 
the Section \ref{sec:reduction} that it cannot be used in order to perform the 
F-theory to M-theory duality. 

\section{Fluxed $S^1$ reduction of the six-dimensional theory}
\label{sec:reduction}
In order to verify and further concretize the six-dimensional
effective theory of subsection~\ref{sec:6dmassiveU(1)} obtained by
compactifying F-theory on $\tX$ one has to take a detour via M-theory.
Therefore, our strategy, as depicted in figures
\ref{fig:resolved_side} and \ref{fig:deformed_side}, is to compactify
the six-dimensional effective theories of subsections
\ref{sec:6dmasslessU(1)} and \ref{sec:6dmassiveU(1)} on a circle and
compare the resulting five-dimensional effective theory with M-theory
reduced on $\hX$ and $\tX$, respectively.  In subsection
\ref{sec:whXreduction} we recall the circle reduction for $\hX$ that
yields two massless $U(1)$s in five dimensions.  For the fibration
$\tX$ with a bi-section, however, it turns out that a circle reduction
alone can never yield the correct match. In fact, we will argue in
subsection \ref{sec:circle-flux} that it is crucial to include
background fluxes for the gauged axion $c$ in \eqref{gauged_U1} in
order to ever be able to match the effective theories.  The effective
theory obtained after circle reduction with fluxes is derived in
subsection \ref{sec:eff_circle-flux} and compared with the effective
theory for $\tX$.  We stress that analyzing classical and one-loop
Chern-Simons terms in the five-dimensional effective theories is
crucial to establish the duality.

\subsection{Massless U(1) on a circle and its M-theory dual} \label{sec:whXreduction}

\begin{wrapfigure}{r}{0.4\textwidth}
\vspace{-0.8cm}
 \begin{center}
  \begin{tikzpicture}
    
    \kbbox[90, 145,north,F_resolved,(0, 0)] {\textbf{F-theory on $\hX$}};
    \node [] () at (-1, 0.6) {Massless sector:};
    \node [] () at (0.2, -0.6) {\begin{tabular}{ll}
     1 & gauge field $\hat A^1$\\
     $H_{U(1)}$ & charged hypers \\
     $H_{neutral}$ & neutral hypers \\
  \end{tabular}};
  
  \kbbox[150, 155,north,F_resolved_S1,(0, -6.5)] {\textbf{F-theory on $\hX \times S^1$}};
    \node [] () at (-1.2, -4.9) {Massless sector:};
    \node [] () at (0.2, -5.8) {\begin{tabular}{ll}
     2 & gauge fields $A^a$\\
     $H_{neutral}$ & neutral hypers \\
  \end{tabular}};
    \node [] () at (-1.2, -7) {Massive sector:};
    \node [] () at (0.0, -8.2) {\begin{tabular}{ll}
     $H_{U(1)}$ & hypers charged\\
     & under $A^1$\\
     \multicolumn{2}{l}{+ KK towers of all fields}
  \end{tabular}};

  \draw [-latex] (F_resolved.south) -> node[above, yshift=-3.5em, xshift=1.7em]{\begin{turn}{270}
 \begin{tabular}{c}compactify\\ on $S^1$\end{tabular}
 \end{turn}}(F_resolved_S1.north);

  \kbbox[70, 160,north,M_resolved,(0, -13.2)] {\textbf{M-theory on $\hX$}};
    \node [] () at (-1.2, -12.9) {Massless sector:};
    \node [] () at (0, -13.7) {\begin{tabular}{ll}
     2 & gauge fields $A^a$ \\
     $H_{neutral}$ & neutral hypers \\
  \end{tabular}};
  
 \draw [-latex] (F_resolved_S1.south) -> node[above, yshift=-4em, xshift=1.7em]{\begin{turn}{270}
 \begin{tabular}{c}integrate out \\ massive states \end{tabular}
 \end{turn}}(M_resolved.north);
 
  \end{tikzpicture}
  \end{center}
  \vspace{-0.5cm}
  \caption{The different theories related to the resolved manifold $\hX$ and their interrelations.}
  \label{fig:resolved_side}
  \vspace{0.7cm}
  \end{wrapfigure}

In this subsection we review the five-dimensional effective action 
obtained by compactifying an F-theory model with one massless $U(1)$ 
on a circle. We also comment on the one-loop effective theory which 
one obtains by entering the Coulomb branch of the five-dimensional theory and 
integrating out all massive modes. This amounts to discussing
the first column of figure \ref{fig:overview}, which we reproduce
in more detail in figure \ref{fig:resolved_side}.

The Kaluza-Klein ansatz for the six-dimensional metric is given by 
\begin{equation} \label{6dmetric_ansatz}
  ds_{(6)}^2 = g_{\mu \nu} dx^\mu dx^\nu  + r^2 (dy  - A^0)^2\ , 
\end{equation}
where $r$ is the radius of the $S^1$ and  
$A^0$ is the Kaluza-Klein vector that will play a crucial role in the 
following. The $U(1)$ vector $A^1$ reduces on a circle 
as 
\begin{equation}
   \hat A^1 = A^1 + \zeta (dy - A^0)\ ,
\end{equation}
with the vector $A^1$ and the scalar $\zeta$ forming the bosonic components 
of a five-dimensional vector multiplet. In addition, there are $T+1$ 
five-dimensional vectors $A^\alpha$ arising from six-dimensional tensors
$\hat B^\alpha$ and $T+1$ scalars $j^\alpha$ satisfying
one constraint $j^\alpha j^\beta \Omega_{\alpha \beta} =1$. 
Note that in this section \ref{sec:reduction} all scalars including 
$c,h^s$ live in a five-dimensional space-time.

Let us next package the reduced fields into five-dimensional vector multiplets
and introduce the five-dimensional theory. To begin with, recall that
the dynamics of the $T+2$ vector multiplets and the graviphoton are entirely specified
in terms of a cubic potential
\begin{equation} \label{cN}
   \mathcal{N} = \tfrac{1}{3!} k_{I J K} M^{I} M^{J} M^{K}\,.
\end{equation}
where  $k_{IJK}$ is a constant symmetric tensor. The potential $\cN$
depends on the real coordinates $M^I$, 
$I=0, \dots, T+2$ and encodes a
real special geometry of $\cN=2$ supergravity. The $M^I$ combine with the vectors $A^I$ of the
theory. However, since the vector in the gravity multiplet is not accompanied
by a scalar degree of freedom, the $M^I$ have to satisfy one constraint.
In fact, the $\cN=2$ scalar field space is identified with the hypersurface $\mathcal{N} \overset{!}{=} 1$.
The gauge coupling function and the metric are obtained by evaluating the second derivative
of $ -\frac12 \log \mathcal{N} $ restricted to the constraint hypersurface.
For completeness, let us give the $M^I$ for the circle reduced setup:
\beq
 M^0= \frac{1}{2} r^{-4/3}\ ,\quad M^1 =2  r^{-4/3} \zeta\ , \quad M^\alpha =2  r^{2/3} (j^\alpha + 2 b^\alpha \zeta^2/r^2)\ .
\eeq
The F-theory reduction with $U(1)$s was carried out in \cite{Grimm:2013oga} and
it was found that the cubic potential takes the form\footnote{We remark that there is an additional
non-polynomial part acting as local counterterms in the five-dimensional action. As it does not take 
part in the match with M-theory, we omit it here and refer to \cite{Bonetti:2011mw,Grimm:2013oga}
for further information.}
\begin{equation} \label{NF}
   \cN^{\rm F} = \frac{1}{2} M^0 \Omega_{\alpha \beta} M^\alpha M^\beta - \frac{1}{2} M^\alpha \Omega_{\alpha \beta} b^\beta M^1 M^1\,.
\end{equation}
It should be stressed that this is only the classical contribution with all charged hypermultiplets retained in
the five-dimensional theory. As we will discuss below, equation \eqref{NF} receives one-loop corrections from
integrating out massive modes, such as the Kaluza-Klein states.

In the following we will mostly focus on the couplings of the vectors $A^I=(A^0,A^1,A^\alpha)$,
as supersymmetry then also determines the vector multiplet couplings of the action. 
In particular, we will discuss the Chern-Simons action for the 
vectors 
\begin{equation} \label{gen_CSterms}
    S_{\rm CS} = -\frac{1}{12} \int_{M^{4,1}} k_{IJK} A^I \wedge F^J \wedge F^K
    - \frac{1}{4} \int_{M^{4,1}} k_I A^I \wedge \tr (\mathcal{R} \wedge \mathcal{R})\ ,
\end{equation}
where $F^I = dA^I$ and $\mathcal{R} $ is the five-dimensional curvature two-form.
Classically, the Chern-Simons coefficients $k_{IJK}$ can be read off from \eqref{NF} as
\begin{equation} \label{k3class}
   k^{\rm class}_{0 \alpha \beta} = \Omega_{\alpha \beta}\ , \qquad k^{\rm class}_{ \alpha 1 1} = - \Omega_{\alpha \beta} b^\beta\ ,
\end{equation}
with all other classical triple couplings vanishing. 
In addition, one finds that $k_I$ is classically given by
\begin{equation} \label{k1class}
    k_\alpha^{\rm class} = \Omega_{\alpha \beta} a^\beta\ ,  \qquad k_0^{\rm class} = 0 \ , \qquad k^{\rm class}_1 = 0 \ ,
\end{equation}
with $a^\alpha$ as in \eqref{SGS}.

In is important to stress that the classical theory with Chern-Simons terms
\eqref{k3class} and \eqref{k1class} cannot 
be successfully compared with the M-theory reduction on the non-singular manifold $\hX$.
To make such a comparison, one first has to move to the five-dimensional Coulomb branch 
by giving the scalar $\zeta$ in the vector multiplet of the six-dimensional extra $U(1)$ a vacuum 
expectation value. Furthermore, one has to integrate out all massive states. 
In general, the mass of a state $\mathbf{w}$ with Kaluza-Klein charge $\hat n$ is
\begin{equation}
   m_{\mathbf{w}}(\hat n) = m^{\mathbf{w}}_{CB} + \hat n\ m_{KK} \,.
\end{equation}
Note that the Coulomb branch mass $m^{\mathbf{w}}_{CB}$ depends on the charges
$\mathbf{w}_i$ of the state $\mathbf{w}$ via $m^{\mathbf{w}}_{CB} = q_i(\mathbf{w}) \zeta^i$.
Integrating out a state causes the Chern-Simons terms of $A \wedge \tr \cR \wedge \cR$
and $A \wedge F \wedge F$ to shift according to~\cite{Bonetti:2013ela}\footnote{The spin-1/2 case was first discussed in \cite{Witten:1996qb}.}
\begin{align}
k_{\Lambda \Sigma \Theta} & \mapsto  k_{\Lambda \Sigma \Theta} 
  + c_{AFF} \  q_{\Lambda} q_{\Sigma} q_{\Theta} \sign(m) \\
  k_{\Lambda} &\mapsto k_{\Lambda} + c_{A\mathcal{R}\mathcal{R}}\ q_{\Lambda} \sign(m) \,,  
\end{align}
respectively, where $c_{AFF}$ and $c_{A\mathcal{R}\mathcal{R}}$ are constants depending on the sort of
state integrated out. They were computed in \cite{Bonetti:2013ela} and are listed
in Table \ref{t:cs_multiplier} in the conventions used here.
\begin{table}
 \centering
 \begin{tabular}{|c|ccc|}
 \hline
  & spin-1/2 fermion & self-dual tensor $B_{\mu \nu}$ & spin-3/2 fermion $\psi_\mu$ \\
  \hline
  $c_{AFF}$ & $\frac{1}{2}$ & $-2$ & $\frac{5}{2}$\\
  $c_{A\mathcal{R}\mathcal{R}}$ & $-1$ & $-8$ & $19$\\
  \hline
 \end{tabular}
  \caption{The different constant multipliers for the shifts of the Chern-Simons terms. Note
  that the individual multipliers may have to be multiplied by $-1$ depending on the
  chirality of the state.}
  \label{t:cs_multiplier}
\end{table}

To avoid clutter in the results for loop-corrections, let us introduce one more bit of
notation, namely
\begin{align} \label{e:lw}
l_{\mathbf{w}} \equiv \Bigg\lfloor \frac{|m_{CB}^{\mathbf{w}}|}{|m_{KK}|}\Bigg\rfloor \,,
\end{align}
the (floored) ratio of Coulomb branch mass and Kaluza-Klein mass of a state $\mathbf{w}$.
We point out that $l_{\mathbf{w}}$ vanishes as long as the zero section of the compactification
manifold is holomorphic. For non-holomorphic zero sections, however, a modified F-theory limit
leads to important additional contributions~\cite{Grimm:2013oga} and in the examples studied in
section \ref{sec:examples} we will encounter such cases.

Keeping the number of vector multiplets general as $V$ for the time being, one then finds
that the loop-corrected Chern-Simons terms are
\begin{align}
  k_0 &= \frac{1}{6} \left(H-V+5T+15\right) + \sum_{\cR} H(\cR)
 \sum_{\mathbf{w} \in \cR} l_{\mathbf{w}} (l_{\mathbf{w}}+1)  \label{k0} \\
 k_1 &=  \sum_{\cR} H(\cR) \sum_{\mathbf{w} \in \cR}
 q_1({\mathbf{w}}) ( 2 l_{\mathbf{w}} + 1) \sign(m^\mathbf{w}_{CB}) \,, \label{k1}
\end{align}
and
\begin{align}
  k_{000} &= \frac{1}{120} \left(H-V-T-3\right) - \frac{1}{4} \sum_{\cR} H(\cR)
 \sum_{\mathbf{w} \in \cR} l_{\mathbf{w}}^2 (l_{\mathbf{w}}+1)^2  \label{k000} \\
 k_{001} &= -\frac{1}{6} \sum_{\cR} H(\cR) \sum_{\mathbf{w} \in \cR}
 q_1({\mathbf{w}}) l_{\mathbf{w}} (l_{\mathbf{w}} + 1) (2 l_{\mathbf{w}} + 1) \sign(m^\mathbf{w}_{CB})  \label{k001} \\
 k_{011} &= -\frac{1}{12} \sum_{\cR} H(\cR) \sum_{\mathbf{w} \in \cR}
 q_1({\mathbf{w}})^2 \left( 1 + 6 l_{\mathbf{w}} ( l_{\mathbf{w}} + 1) \right)  \label{k011} \\
 k_{111} &= -\frac{1}{2} \sum_{\cR} H(\cR) \sum_{\mathbf{w} \in \cR}
 q_1({\mathbf{w}})^3 ( 2 l_{\mathbf{w}} + 1) \sign(m^\mathbf{w}_{CB})  \label{k111}
\end{align}
where, just as above, we have denoted the charge of the state $\mathbf{w}$
under the six-dimensional $U(1)$ by $q_1(\mathbf{w})$ and sum over all matter representations
$\cR$ that are present in our theory.
Note that since we are considering
purely Abelian models, all representations $\cR$ are
 one-dimensional and therefore contain only a single weight $\mathbf{w}$.

Finally, let us remark on how to use the loop-corrected Chern-Simons terms
in order to compute the matter spectra of the associated F-theory model.
As first explored in \cite{Grimm:2011fx} and later refined in \cite{Grimm:2013oga},
one can make an ansatz for the matter spectrum, keeping the multiplicities
general. Such an ansatz can for example be
based on the curves found in the (relative) Mori cone of the Calabi-Yau, or,
torically, the split induced by the top of the 
compactification manifold \cite{Braun:2013yti, Braun:2013nqa}.
Next, one uses that the matching of the M-theory and F-theory low-energy effective
actions implies that the loop-corrected Chern-Simons terms must be given
by simple topological intersection numbers in the M-theory geometry:
\begin{align}
 k_{IJK} = D_I \cdot D_J \cdot D_K
\end{align}
Here, just as before, the $D_I$ are the divisors corresponding to the fields $A^I$
in the usual manner.

For an explicit compactification manifold $\hX$, we can therefore simply compute
what the loop-corrected Chern-Simons terms of the five-dimensional theory must be by
doing intersection theory on $\hX$. Demanding
that they match the formulas in \eqref{k0} - \eqref{k111} for the chosen ansatz
one hence obtains a system of linear equations for the matter multiplicities.
For all known examples in the literature, this system of equations
has a unique solution.

\subsection{Background flux and the M-theory to F-theory limit for multi-sections} \label{sec:circle-flux}

In this subsection we argue that a simple circle reduction is not sufficient 
when considering F-theory on the Calabi-Yau threefold $\tX$ with a multi-section.
In order to do that we consider M-theory on $\tX$ and 
dualize the setup step by step to obtain a Type IIB compactification. 

To begin, we must consider the different structure of the Calabi-Yau metric in the case that the elliptically fibered space has,
or does not have, a section. Let us denote by $u^i$ the local (complex) coordinates on the base $B_2$ of $\tX$
and by $(x,y)$ local coordinates on the torus fiber. In the case that the fibration admits a section, it is possible
to describe the base $B_2$ as a complex (algebraic) hypersurface within $\tX$ given locally by a defining equation,
$f(x,y,u)=0$. This realization of $B_2$ as a hypersurface (in fact sub-manifold) of $\tX$ makes it possible to use 
geodesics to define coordinates normal to $B_2$ within $\tX$ consistently for each coordinate patch in $B_2$, and
as a result the $3$-fold metric takes a complex, K\"ahler version of Gaussian normal form \cite{smyth, yano}.
That is, the metric can be made block-diagonal with respect to the fiber/base with
$g_{I5}=g_{I6}=0$ for $I=1,\ldots 4$ denoting base directions and $5,6$ fiber directions. 

By contrast, it was noted in \cite{Witten:1996bn} that in the case that $\tX$ has multi-sections only, the base is no longer a submanifold of $\tX$ 
and no such hypersurface description exists. As a result, there must exist {\it some} coordinate patch in $B_2$ for which the diagonalization 
described above fails and $g_{I5}$ and/or $g_{I6}\neq 0$. Let us consider such a patch and over it, take a semi-flat approximation to 
the Calabi-Yau metric \cite{Greene:1989ya,Strominger:1996it,Moss}. Away from any singular fibers the metric takes the local form
\beq \label{metric_withoutsection}
   ds^2(\tX) = g_{i \bar \jmath} \, du^i d\bar u^{\bar \jmath} + \frac{v^0}{\text{Im} \tau} | X - \tau Y|^2 \ ,  
\eeq
where at each point of $B_2$ one parametrizes the complex structure of the torus fiber by $\tau(u)$ and $v^0$ is the overall area of the $T^2$ fiber,
which is constant over the base.
The presence of off-diagonal (fiber/base) metric components are 
parametrized here by vectors $(\tilde X,\tilde Y)$ on $B_2$ in 
\beq
   X = dx + \tilde X \ , \qquad Y = dy  + \tilde Y\ , \qquad K = \tilde X - \tau \tilde Y\ ,
\eeq 
where we have introduced a complex vector $K$ on $B_2$ in order to re-write the metric in complex coordinates. Defining $z=x -\tau y$, \eref{metric_withoutsection} takes the form
\beq
 ds^2(\tX) = g_{i \bar \jmath} \, du^i d\bar u^{\bar \jmath} + \frac{v^0}{\text{Im} \tau} |dz - \frac{\text{Im} z\, d\tau}{\text{Im} \tau}+ K|^2 \ .
\eeq
We locally define on $\tX$ the two-form 
\beq
 \omega_0 = \frac{1}{\text{Im}{\tau}}(dz - \frac{\text{Im} z \, d\tau}{\text{Im} \tau}+K)\wedge (d{\bar z}- \frac{\text{Im}z\, d{\bar \tau}}{\text{Im} \tau}+{\bar K}) = 2 Y \wedge X 
\eeq
In terms of $\omega_0$ the globally defined two-form on $\tX$ is given by $J = J_{\rm base} + v^0 \omega_0$.
If $K$ is a $(1,0)$ form then $J$ is of type $(1,1)$ and we find compatibility of \eref{metric_withoutsection} with the complex structure
\cite{2002math2282C}. Using that $\tau$ is holomorphic in the base coordinates it follows that $d(K/\text{Im}\tau)$ and $d(\bar{K}/\text{Im}\tau)$ 
are both $(1,1)$ forms. Together with the fact that
 \beq
 \frac{i(K-\bar{K})}{2\text{Im}\tau}=\tilde Y \ , \qquad \frac{i({\bar{\tau}}K-\tau\bar{K})}{2\text{Im}\tau}=\tilde X
 \eeq
 we obtain finally that $\langle d\tilde X \rangle$ and $\langle d\tilde Y \rangle$ are $(1,1)$ forms.
 In the following we will consider the case that 
 \beq \label{dX=nO}
   \langle d\tilde X \rangle = - n \tilde \omega \ , \qquad \langle d\tilde Y \rangle = 0\ ,
 \eeq
where $\tilde \omega$ is an appropriately normalized (1,1) form on $B_2$, 
which has to be identified with the form appearing in \eqref{C2expansion}. The ansatz 
\eqref{dX=nO} implies the presence of exactly one gauged axion $c$ and has 
to be generalized accordingly for more involved situations. In this simplest setup, however,
$\langle d\tilde Y \rangle$ has to vanish for the consistency of the effective theory. 

In the following we consider M-theory on the space \eqref{metric_withoutsection} and
perform the M-theory to F-theory limit. 
The eleven-dimensional metric and M-theory three-form are expanded as
\begin{align}
  ds^{2}_{11} = ds^2_5 + ds^2(\tX) \,, \qquad C_3^M = B_2 \wedge X + C_2 \wedge Y + \frac{1}{2} A^0 \wedge \omega_0 + \ldots \,,  \label{CMexp}
\end{align}
where the dots indicate the expansion into further harmonic (1,1) forms of $\tX$ irrelevant to 
the present discussion. We also expand $B_2 = b \tilde \omega$ and $C_2 = c \tilde \omega$
and compute 
\beq
   dC_3^M = db \wedge X \wedge \tilde \omega + b\, \tilde \omega^2 + (dc + n A^0 )\wedge Y  \wedge \tilde \omega  +  \frac{1}{2} F^0 \wedge \omega_0 + \ldots\ ,
\eeq
where we have used $d \omega_0 =  2 n\,  Y \wedge \tilde \omega$. We note that the non-trivial background 
$\langle d\tilde X \rangle$ implies that the axion $c$ is gauged by the vector $A^0$. Following the 
M-theory to F-theory duality, which we discuss next, one finds that with the expansion \eqref{CMexp} the vector 
$A^0$ maps precisely to the Kaluza-Klein vector of 
the reduction from six to five dimensions. 

Due to the presence of non-trivial 
$\tilde X,\tilde Y$ in \eqref{metric_withoutsection} the standard M-theory to F-theory limit is 
modified (see \cite{Denef:2008wq} for a review). 
To fix an $SL(2,\mathbb{Z})$ frame, let us pick an A-cycle and a B-cycle
of the genus-one fiber with local coordinates $x$ and $y$, respectively. 
In order to perform the duality we first go from M-theory to Type IIA
by splitting the metric with respect to the A-cycle according to
\begin{equation}
   ds_M^2 = e^{4 \phi_{{\rm IIA}} /3} (dx + C^{\rm IIA}_1)^2 + e^{-2 \phi_{\rm IIA}/3} ds_{\rm IIA}^2 \,.
\end{equation}
Comparing with \eqref{metric_withoutsection} one finds the Type IIA R-R one-form $ C^{\rm IIA}_{1}$ and metric $ds_{\rm IIA}^2$ to be 
\begin{align}
   C^{\rm IIA}_{1} = \R\, \tau\,  dy + \R\, K\,, \qquad
   ds_{\rm IIA}^2 = \sqrt{\frac{v^0}{\I \tau}}\Big(\frac{v^0}{\I \tau} (\I \, \tau \, dy +  \I\, K)^2 
   +g_{i \bar \jmath} \, du^i d\bar{u}^{\bar \jmath}   \Big)
\end{align}
with $e^{4 \phi_{\rm IIA}/3} = \frac{v}{\text{Im}\tau}$.
Using the T-duality rules along the 
B-cycle one encounters non-trivial NS-NS and R-R two-forms
\begin{equation} \label{B2C2}
     C^{\rm IIB}_2 = C_2 + \tilde X \wedge dy \ , \qquad B^{\rm IIB}_2  = B_2 + \tilde Y \wedge dy \ .
 \end{equation}
The presence of non-trivial $C^{\rm IIB}_2$ and $B^{\rm IIB}_2$ in \eqref{B2C2} implies 
that the F-theory reduction should include three-form fluxes
\beq \label{G3dy}
    F_3 =  \langle dC^{\rm IIB}_2 \rangle = -   n\, \tilde \omega \wedge dy \ .
\eeq 
We stress that this flux has one leg around the circle used to compactify
six to five dimensions. 

Let us now make contact with the discussion of subsection \ref{sec:6dmassiveU(1)}. 
After decompactifying the T-dualized Type IIB circle the scalars $c,b$ are
lifted to proper six-dimensional scalars. One can then reinterpret that flux 
\eqref{G3dy}. Compactifying the six-dimensional theory on a circle the 
flux $n$ can be understood as a background of $dc$ given by
\beq
   \int_{S^1} \langle dc \rangle = n\ .
\eeq
This implies that the standard circle reduction has to include this non-trivial background 
and we will explicitly perform this modified computation in the next subsection.

\subsection{Fluxed circle reduction and M-theory comparison} \label{sec:eff_circle-flux}

\begin{wrapfigure}{r}{0.4\textwidth}
\vspace{-.3cm}

\centering
  \begin{tikzpicture}  
  \kbboxred[90, 170,north,F_deformed,(10, -0.5)] {\textbf{F-theory on $\tX$}};
    \node [] () at (8.6, 0.1) {Massless sector:};
    \node [] () at (10.2, -0.7) {\begin{tabular}{ll}
     $H_{U(1)}-1$ & charged hypers \\
     $H_{neutral}$ & neutral hypers \\
  \end{tabular}};
    \node [] () at (9.4, -1.8) {$1$ massive gauge field $A^1$};

  \kbbox[180, 185,north,F_deformed_S1,(10, -6.5)] {\textbf{F-theory on $\tX \times S^1$}};
    \node [] () at (8.3, -4.3) {Massless sector:};
    \node [] () at (10.2, -5.5) {\begin{tabular}{ll}
     1 & gauge field $\widetilde{A}^0$\\
     $H_{neutral} +\delta -1$ & hypers neutral \\
     & under $\widetilde{A}^0$
  \end{tabular}};
    \node [] () at (8.3, -6.9) {Massive sector:};
    \node [] () at (9.8, -8.3) {\begin{tabular}{ll}
     $1$ & gauge field $\widetilde{A}^1$\\
     $H_{U(1)} - \delta$ & hypers charged\\
     & under $\widetilde{A}^1$ \\
     \multicolumn{2}{l}{+ KK towers of all fields} 
  \end{tabular}};
    
   \draw [-latex] (F_deformed.south) -> node[above, yshift=-2em, xshift=4em]{
   \begin{tabular}{c}compactify on\\$S^1$ with flux\end{tabular}
   }(F_deformed_S1.north);
  \kbbox[75, 185,north,M_deformed,(10, -12.5)] {\textbf{M-theory on $\tX$}};
    \node [] () at (8.3, -12.3) {Massless sector:};
    \node [] () at (10.1, -13.1) {\begin{tabular}{ll}
     1 & gauge field $\widetilde{A}^0$ \\
     $H_{neutral} + \delta - 1$ & neutral hypers \\
  \end{tabular}};
  \draw [-latex] (F_deformed_S1.south) -> node[above, yshift=-2em, xshift=4em]{
  \begin{tabular}{c}integrate out \\ massive states \end{tabular}
  }(M_deformed.north);
  \end{tikzpicture}
  
  \caption{The different theories related to the deformed manifold $\tX$ and their interrelations.}
  \label{fig:deformed_side}
  
  \end{wrapfigure}

Having motivated the inclusion of circle fluxes we are now in the position 
to compute the five-dimensional effective theory, that is,
we proceed by discussing the second column of figure \ref{fig:overview},
reproduced in figure \ref{fig:deformed_side} with the relevant matter
spectra included. In performing this reduction we include the circle fluxes
\beq
   \int_{S^1} \langle dc \rangle = n \ . 
\eeq
Using the background metric \eqref{6dmetric_ansatz} this 
implies that the kinetic term of the axion $c$ reduces as 
\beq
   \cL_{c} =  G_{cc} |\mathcal{\hat D} c |^2  = G_{cc} |\mathcal{D} c|^2\ ,
\eeq
where $G_{cc}$ is the metric for the field $c$.
In other words, the six-dimensional invariant derivative of the 
axion $c$ given in \eqref{gauged_U1} is replaced by 
\beq
   \mathcal{D} c = dc + m A^1 + n A^0\ , 
\eeq
We stress that this modification only appears in the five-dimensional 
effective theory and mixes the reduced $U(1)$ vector $A^1$ with 
the Kaluza-Klein vector $A^0$.

This implies that after absorbing the axion $c$
the mass term in the five-dimensional theory reads
\beq \label{mass_term}
   \mathcal{L}_{\rm mass} =    G_{cc} |m A^1 + n A^0|^2\ ,
\eeq
To evaluate the effective theory for the massless degrees of 
freedom only, we therefore first have to chose an appropriate 
basis of one massless vector field $\tA^0$ and one 
massive vector field $\tA^1$. 

Starting with the two gauge fields $A^0$ and $A^1$, the most general transformation
to a new basis of gauge fields $\tA^0$ and $\tA^1$ can be expressed as
\begin{align} \label{e:basis_change}
 \tA^i = \frac{1}{a^2+b^2}\, N^i{}_j \, A^j\,,
 \qquad N^i{}_j = \begin{pmatrix} b & -a \\ a & b \end{pmatrix}\,.
\end{align}
Note that the orthogonality of the columns of $N^i{}_j$ guarantees
that the kinetic terms of $\tA^i$ remain diagonal under the transformation
if they are already diagonal before. In the following we 
like to identify $\tA^1$ with the massive $U(1)$ with mass term \eqref{mass_term}. 
This implies that $a, b$ in \eqref{e:basis_change} 
are identified to be
\beq \label{abindent}
   a = n \ , \qquad b= m\ .
\eeq
We also need to transform the charges $q_j(\mathbf{w})$ under 
the $A^i$ of a state $\mathbf{w}$. The transformation \eqref{e:basis_change} 
introduces new charges $\tq_i$ as
\begin{align} \label{e:basis_change_charge}
 \tq_i = q_j \, (N^T)^j{}_i\,.
\end{align}

To compare the fluxed circle reduction to the M-theory reduction 
on $\tX$ we thus rotate into the new basis $\tA^i$ and 
then drop the couplings of the massive gauge field $\tA^1$. 
As in section \ref{sec:whXreduction} we have to consistently integrate out all massive 
modes. The way to check that the reduction of the proposed six-dimensional F-theory action 
indeed matches we will identify the five-dimensional Chern-Simons terms. 
We note that the constant couplings $k_{IJK}$ and $k_{I}$ in \eqref{gen_CSterms}
transform under the basis change \eqref{e:basis_change} as
\begin{align} \label{tk-def}
 &\tk_{ijk} = k_{abc} \, (N^T)^a{}_i \, (N^T)^b{}_j\, (N^T)^c{}_k\ , \qquad 
 &&\tk_{ij\alpha} = k_{ab\alpha} \, (N^T)^a{}_i \, (N^T)^b{}_j\ , & \\
 &\tk_{i\alpha\beta} = k_{a\alpha \beta} \, (N^T)^a{}_i \ , \qquad &&\tk_{i} = k_{a} \, (N^T)^a{}_i & \nn
\end{align}
with $\tk_{\alpha \beta \gamma}=k_{\alpha \beta \gamma}=0$ and $\tk_{\alpha} = k_\alpha$ as above. 
Using these expressions together with \eqref{e:basis_change}, \eqref{abindent}, \eqref{k3class} and \eqref{k1class} we 
find the non-vanishing classical Chern-Simons terms for the massless five-dimensional 
gauge fields $(\tA^0,A^\alpha)$ to be
\bea \label{e:cs_deformed_classical}
  \tk^{\rm class}_{00\alpha} &=& -n^2\, \Omega_{\alpha \beta} b^\beta   \ ,\qquad \qquad \tk^{\rm class}_{0 \alpha \beta} = m \Omega_{\alpha \beta} \ ,\\
  \tk^{\rm class}_{\alpha} & = & \Omega_{\alpha \beta} a^\beta \ .
\eea
Let us stress that $ \tk^{\rm class}_{00\alpha} $ is non-zero and depends on the 
classical coupling of the extra U(1). In contrast, if the $\tA^0$ is only the Kaluza-Klein 
vector, one recalls from \eqref{k3class} that $k_{00\alpha} = 0$. The latter is indeed true for 
all models with multiple sections considered in the literature so far. 
Crucially, in the examples with multi-section this coupling no longer vanishes as 
we discuss below and show for specific examples in section \ref{sec:examples}.

The Chern-Simons terms induced by integrating out the massive states 
at one loop level are obtained from \eqref{tk-def} using 
\eqref{k000}-\eqref{k1}. 
For the triple coupling one finds for the massless gauge field $\tA^0$ that
\begin{align} 
  \tk_{000} &= k_{000} m^3 - 3 k_{001} n m^2
 + 3 k_{011} n^2 m - k_{111} n^3\,\\
 &= \frac{m^3}{120} \left(H-V-T-3\right) \nonumber \\
 & \quad + \frac{1}{4} \sum_{\cR} H(\cR) \sum_{\mathbf{w} \in \cR} \Big(
 - m^3  l_{\mathbf{w}}^2 (l_{\mathbf{w}}+1)^2  \nonumber \\[-.4cm]
 & \qquad \qquad \qquad  \qquad \qquad + 2 n m^2
 q_1({\mathbf{w}}) l_{\mathbf{w}} (l_{\mathbf{w}} + 1) (2 l_{\mathbf{w}} + 1) \sign(\mathbf{w}) \nonumber \\
 & \qquad \qquad \qquad \qquad \qquad - n^2 m  q_1({\mathbf{w}})^2
  \left( 1 + 6 l_{\mathbf{w}} ( l_{\mathbf{w}} + 1) \right) \nonumber \\
 & \qquad \qquad \qquad \qquad \qquad 
 + 2 n^3 q_1({\mathbf{w}})^3 ( 2 l_{\mathbf{w}} + 1) \sign(\mathbf{w}) \Big)\,. \label{1-loop_tildek000}
\end{align}
Furthermore, one finds the one-loop contribution to $k_I$ to be
\begin{align} \label{1-loop_tildek}
 \tk_0&= k_0 m - k_1 n \nonumber \\
 & =  \frac{m}{6} \left(H-V+5T+15\right) \nonumber \\
 & \quad + \sum_{\cR} H(\cR) 
 \sum_{\mathbf{w} \in \cR} 
 \Big( m l_{\mathbf{w}} (l_{\mathbf{w}}+1) 
 - n q_1({\mathbf{w}}) ( 2 l_{\mathbf{w}} + 1) \sign(\mathbf{w}) \Big)\,. 
 \end{align}

 Having presented the field theory result for the Chern-Simons 
 terms obtained by integrating out all massive modes, we are now in the 
 position to compare this with the reduction on $\tX$. To try and understand
 the above discussion from a different angle, let us consider the fiber geometry
 of a bi-section for a moment. By definition, a bi-section cuts
 out two different points over a generic point in the base manifold.
 Let us call these points $P$ and $Q$. \emph{Locally}, the
 bi-section is therefore indistinguishable from the sum of two separate sections
 cutting out $P$ and $Q$, respectively. In a given patch, one could therefore
 try and define divisors $V(P)$ and $V(Q)$ and follow the usual procedure
 of applying the Shioda map \cite{shioda1989mordell, shioda1990mordell} 
 to obtain a suitable set of massless gauge fields.
 Choosing $V(P)$ as the zero section, one would thus obtain the two ``local divisors''
 \begin{align}
  D_0 = V(P)\,, \qquad D_1 = \lambda \left( V(Q) - V(P) \right)
 \end{align}
 up to some irrelevant vertical parts, where $\lambda$ is an arbitrary normalization constant.
 However, since we have a bi-section, \emph{globally} the two points $P$ and $Q$
 undergo monodromies and the only well-defined quantity is the divisor $V(P) + V(Q)$.
 Consequently, as the massless $U(1)$ gauge field corresponds to the bi-section, its
 associated divisor must satisfy
 \begin{align} \label{eq:surviving_divisor}
  \widetilde{D}_0 \sim 2 \lambda D_0 + D_1\,,
 \end{align}
 where the proportionality constant is just another normalization factor that we can choose
 arbitrarily. Comparing \eqref{eq:surviving_divisor} to $\widetilde{A}^0$,
 one hence finds
 \begin{align}
  m = 2 \lambda\,, \qquad n = -1\,.
 \end{align}
This geometric argument therefore implies that the fluxes present in the circle reduction
are in fact fixed uniquely up to physically irrelevant rescalings of the massless $U(1)$
gauge field.

\section{Examples: transitions removing the section}
\label{sec:examples}

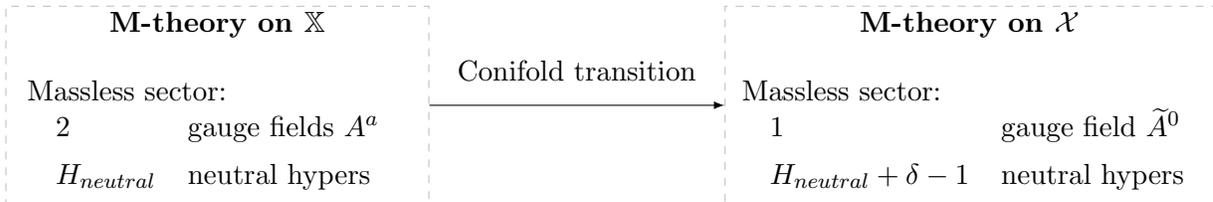
\begin{figure}[h]
\centering
  \begin{tikzpicture}

  \kbbox[75, 160,north,M_resolved,(0, -16.5)] {\textbf{M-theory on $\hX$}};
    \node [] () at (-1.2, -16.3) {Massless sector:};
    \node [] () at (0, -17.1) {\begin{tabular}{ll}
     2 & gauge fields $A^a$ \\
     $H_{neutral}$ & neutral hypers \\
  \end{tabular}};
  
  \kbbox[75, 185,north,M_deformed,(10, -16.5)] {\textbf{M-theory on $\tX$}};
    \node [] () at (8.3, -16.3) {Massless sector:};
    \node [] () at (10.1, -17.1) {\begin{tabular}{ll}
     1 & gauge field $\widetilde{A}^0$ \\
     $H_{neutral} + \delta - 1$ & neutral hypers \\
  \end{tabular}};
  \draw [-latex] (M_resolved.east) -> node[above, yshift=0em, xshift=0em]{\begin{turn}{0}
 \begin{tabular}{c}Conifold transition\end{tabular}
 \end{turn}}
 (M_deformed.west);
  
  \end{tikzpicture}
  \caption{The two theories obtained by compactifying M-theory on $\hX$ and $\tX$, respectively,
  are connected by a conifold transition.}
  \label{fig:conifold_transition}
  \end{figure}

The discussion so far has been general. We now illustrate how the
physics works in a particularly transparent set of examples. These are
given by pairs of Calabi-Yau threefolds $(\hX,\tX)$ related by a
conifold transition, where $\hX$ has two independent sections
and $\tX$ has no section, but rather a multi-section.
Our discussion begins in subsection \ref{ss:pairs_base_independent}
by keeping the treatment of the $(\hX, \tX)$ pairs independent of 
the base manifolds. In subsection \ref{ss:conifolds} we review some well-known
facts about the physics of conifold transition, before
we proceed in subsection \ref{ss:pairs_base_p2} by constructing explicit Calabi-Yau
manifolds with base manifold $\mathbb{P}^2$.
Finally, we evaluate the Chern-Simons terms of some of the specific 
examples in subsection \ref{ss:cs_terms} and give a general argument
explaining why they have to match.
In figure \ref{fig:conifold_transition} we give a pictorial description
of the essential physical process studied in the following subsections.

\subsection{Constructing $(\hX, \tX)$ pairs with general base manifold} \label{ss:pairs_base_independent}

The basic observation allowing us to construct large numbers of such
pairs is that there is a natural conifold transition implicit in most
recent constructions of spaces with two sections. As described in
\cite{Morrison:2012ei}, for example, the generic model with two
sections is obtained by taking a Calabi-Yau hypersurface in
$\wh{\bP^{1,1,2}}$. Let us parametrize $\wh{\bP^{1,1,2}}$ by the
coordinates
\begin{align}
  \begin{array}{c|cccc}
    & y_1 & y_2 & w & t\\
    \hline
    \bC^*_1 & 1 & 1 & 2 & 0\\
    \bC^*_2 & 0 & 0 & 1 & 1
  \end{array}
\end{align}
We blow-up the $\bZ_2$ singularity in the fiber to have a nicer
ambient space, and to be able to realize torically the Cartan divisor
in some of the examples below. The Stanley-Reisner ideal (\emph{SRI}
in what follows) is generated by $\langle y_1y_2, wt\rangle$. The
generic Calabi-Yau hypersurface is a degree $(4,2)$ hypersurface in
these coordinates, which we parametrize as
\begin{align}
  \label{eq:quartic}
  g w^2 + wt P(y_1,y_2) + t^2 Q(y_1,y_2) = 0\, ,
\end{align}
with $P(y_1,y_2)$ a quadratic function in $y_i$
\begin{align}
  \label{eq:P}
  P(y_1,y_2) = \alpha y_1^2 + \beta y_1y_2 + f y_2^2
\end{align}
and $Q(y_1,y_2)$ a quartic
\begin{align}
  Q = y_1(by_1^3 + cy_1^2y_2 + d y_1y_2^2 + ey_2^3) + ay_2^4 \equiv
  y_1 Q'(y_1,y_2) + ay_2^4\, .
\end{align}
Since the elliptic fiber will be fibered over a base, $g$ and the
coefficients of $P, Q$ will be sections of appropriate degree in the
coordinates of the base (we will study some explicit examples
below).\footnote{The models constructed in \cite{Morrison:2012ei}
  correspond to taking $g=1$, which imposes some restrictions on the
  allowed fibrations. We do not impose such restriction.} In order to
have two sections, we set $a=0$, so $Q$ takes the form
\begin{align}
  Q = y_1(by_1^3 + cy_1^2y_2 + d y_1y_2^2 + ey_2^3)= y_1 Q'(y_1,y_2)\, .
\end{align}
The restricted Calabi-Yau equation becomes
\begin{align}
  \label{eq:restricted-quartic}
  \phi\equiv gw^2 + wt P(y_1, y_2) + t^2 y_1 Q'(y_1, y_2) = 0\, .
\end{align}
When the coefficients are chosen in this way, there are two sections
of~\eqref{eq:restricted-quartic} that can easily be found. Take $y_1=0$. Since
$y_1y_2$ belongs to the SRI of $\wh{\bP^{1,1,2}}$, we can set
$y_2=1$. We end up with
\begin{align}
  \label{eq:quartic-restricted-sections}
  w(gw + t f) =0
\end{align}
where $f$ is the coefficient of $y_2^2$ in $\sigma_0$ (again a section of
some line bundle on the base, in general). We thus find a first
section at $w=0$ (we can then set $t=1$ using $\bC^*_2$), and a second
section at $gw=-tf$. For generic choices of $g,f$ and at generic
points of the base, this equation has a unique solution, giving a
second section, but at the zeroes of $g,f$ it will behave in
interesting ways.

\paragraph{Singularities.} The
hypersurface~\eqref{eq:restricted-quartic} will be singular when
$\phi=d\phi=0$. It is easy to check that solutions of this set of
equations exist for $w=y_1=e=f=0$. For two-dimensional bases of the
fibration, $e=f=0$ generically has a set of solutions given by
points. Close to one such zero, for generic values of the
coefficients, equation~\eqref{eq:restricted-quartic} becomes
\begin{align}
  \lambda_1 w^2 + \lambda_2 wf + \lambda_3 w y_1 + \lambda_4 y_1^2 +
  \lambda_4 y_1 e = 0
\end{align}
where $\lambda_i$ are constants,\footnote{These constants can be
  easily read from~\eqref{eq:restricted-quartic}, but we only need
  that they are non-vanishing constants.} and one should see
$w,y_1,f,e$ as local variables for a $\bC^4$ neighborhood of the
singularity in the ambient space. Generically this is a non-degenerate
quadratic form on the ambient space variables, defining locally a
conifold singularity. For later reference, note that the number of
such singularities is given by the number of points in $e=f=0$, or
slightly more formally by the intersection of the homology classes of
the divisors $[e]\cdot [f]$ on the base. Associated with these
singularities there will be massless hypermultiplets coming from
wrapped M2 branes, which will be the essential states in our
discussion.

\paragraph{Deformation.} Since the singularities are conifolds, we
expect that there are two ways of smoothing out the singularities. The
first is by deformation, i.e. changing the Calabi-Yau
equation~\eqref{eq:restricted-quartic}. Our only option is to consider
deformations away from $a=0$. This indeed modifies the analysis above
in that a singularity would require $a=f=e=0$, but for non-vanishing
$a$ and a two-dimensional base there is generically no solution to
this system (by simple dimension counting), so there is no singularity
anymore. An important observation for our purposes below is that under
this deformation the two sections no longer exist independently, but
they rather recombine into a unique global object. Setting $y_1=0$
in~\eqref{eq:quartic} gives
\begin{align}
  gw^2 + wtf + at^2 = 0 \, ,
\end{align}
which no longer factorizes globally. The two sections above still
exist locally and can be found by solving for $w$, but there is a
$\bZ_2$ monodromy coming from going around zeros of the discriminant
$t^2(f^2 - 4ag)$, which exchanges the two roots. This is thus a case
with a bi-section, but no section. In the examples below the
non-existence of a section can also be easily verified using Oguiso's
criteria \cite{Oguiso,Morrison:1996na}, we collect some of the
relevant details in appendix~\ref{sec:Oguiso}. All in all, this gives
the first element of our pair, the deformed Calabi-Yau threefold
$\tX$.

\paragraph{Resolution.} On the other hand, one can do a blow-up of the
conifold in order to desingularize the geometry. A simple toric way of
achieving this is by blowing up the $y_1=w=0$ point, which is the
point of intersection of the conifolds with the fiber, as done in
\cite{Morrison:2012ei}. More concretely, we replace the fiber by the
following GLSM:
\begin{align}
  \label{eq:resolved-GLSM}
  \begin{array}{c|ccccc}
    & y_1 & y_2 & w & t & s\\
    \hline
    \bC^*_1 & 1 & 1 & 2 & 0 & 0\\
    \bC^*_2 & 0 & 0 & 1 & 1 & 0\\
    \bC^*_3 & 1 & 0 & 1 & 0 & -1
  \end{array}
\end{align}
The new Stanley-Reisner ideal is given by $\langle wy_1, wt, st, sy_2,
y_1y_2\rangle$. Notice in particular that $w=y_1=0$ does not belong to
the ambient space anymore. The Calabi-Yau hypersurface in this space
is of degree $(4,2,1)$ and can be parametrized, matching with the
proper transform of~\eqref{eq:restricted-quartic}, by
\begin{align}
  \label{eq:resolved-quartic}
  \wt{\phi} \equiv gw^2s + wt P(sy_1, y_2) + t^2 y_1 Q'(sy_1, y_2) = 0\, .
\end{align}
The sections transform naturally under the blow-up. In particular, the
$w=y_1=0$ section transforms to $s=0$. Setting $s=0$
in~\eqref{eq:resolved-quartic}, and setting $t=y_2=1$ since they
cannot vanish when $s=0$, one gets
\begin{align}
  wf + y_1e = 0
\end{align}
so this section maps to $(y_1, y_2, w, t, s)=(-f,1,e,1,0)$. Let us
denote this section by $\sigma_0$. We will take it to be our zero
section, parametrizing the F-theory limit.

The other section is given by $y_1=0$. Plugging this
into~\eqref{eq:resolved-quartic}, and setting $w=y_2=1$, one gets
\begin{align}
  gs + tf = 0\, .
\end{align}
We thus find a second section at $(y_1,y_2,w,t,s)=(0,1,1,-g,f)$, which
we denote by $\sigma$. We think of this section as generating a $U(1)$
symmetry in the six-dimensional theory obtained by putting F-theory on
$\hX$, choosing $\sigma_0$ as the zero section.

So, as expected, deformation does not recombine the sections, but
rather we stay with two independent sections of the fibration.

It is also not hard to see that the resulting space is generically
non-singular, as one may have expected from the fact that we are
considering the most general equation over the blown-up
fiber. We denote the resulting space by $\hX$.

\paragraph{Holomorphy of the sections.} Looking at the sections we
just found, we see that they are ill-defined over some points in the
base. In particular, $\sigma_0$ is ill-defined over $f=e=0$, since
over these points $\sigma_0$ would be $(0,1,0,1,0)$, but $y_1w$ is in
the Stanley-Reisner ideal. Similarly, $\sigma$ becomes ill-defined
over $g=f=0$, since $st$ is in the Stanley-Reisner ideal. This is a
hallmark of rationality of the sections (as opposed to holomorphy):
the sections are not given by a single point in the fiber everywhere,
but over some subspaces (where $\sigma_0$ and $\sigma$ becomes
ill-defined in our examples) they wrap components of the fiber.

It is not hard to be more explicit about the behavior of these
sections on the problematic points. Setting $f=e=0$, and $s=0$, the
Calabi-Yau equation~\eqref{eq:resolved-quartic} becomes identically
satisfied, so the section at this point jumps in dimension. Similarly
for $\sigma$, since at $y_1=f=g=0$~\eqref{eq:resolved-quartic} is
identically satisfied, so $\sigma$ again jumps in dimension at these
points.

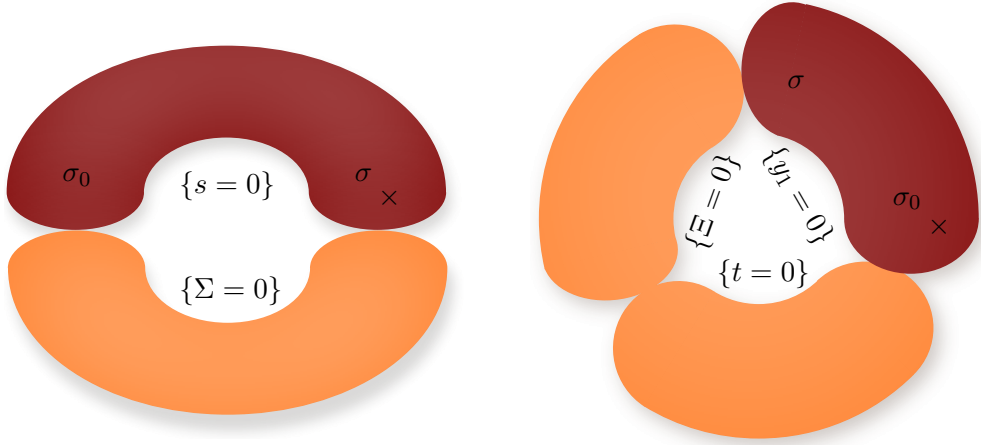
\begin{figure}
\centering
\setlength{\unitlength}{0.1\textwidth} 
\begin{picture}(8.5,3) 
\linethickness{2pt}
\put(0,0){\includegraphics[width=0.4\textwidth]{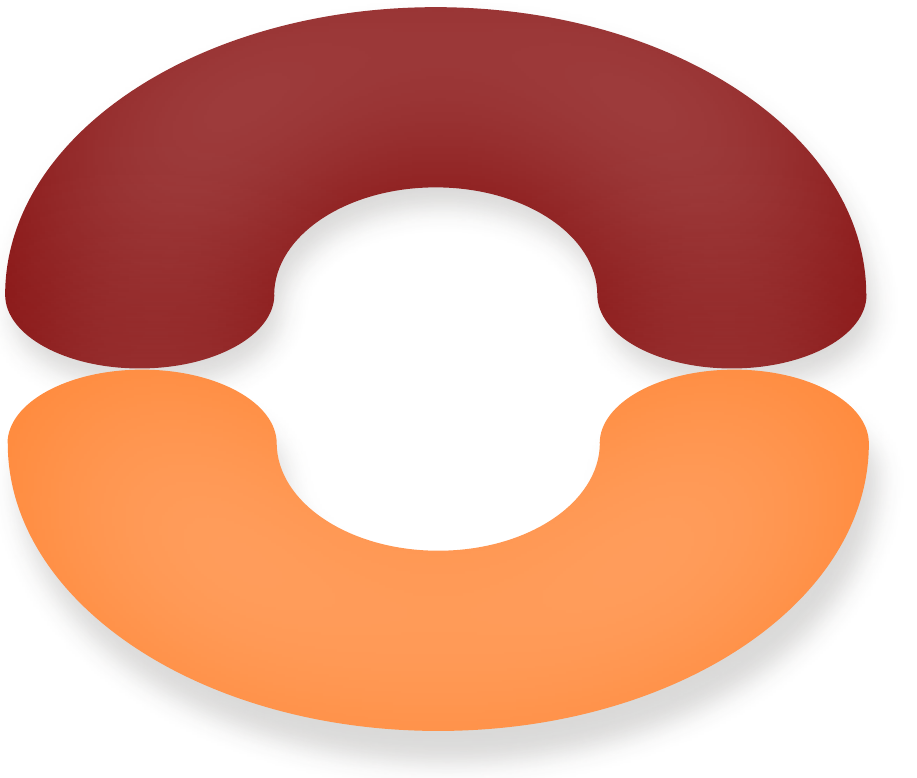}}
\put(4.5,-.2){\includegraphics[width=0.4\textwidth]{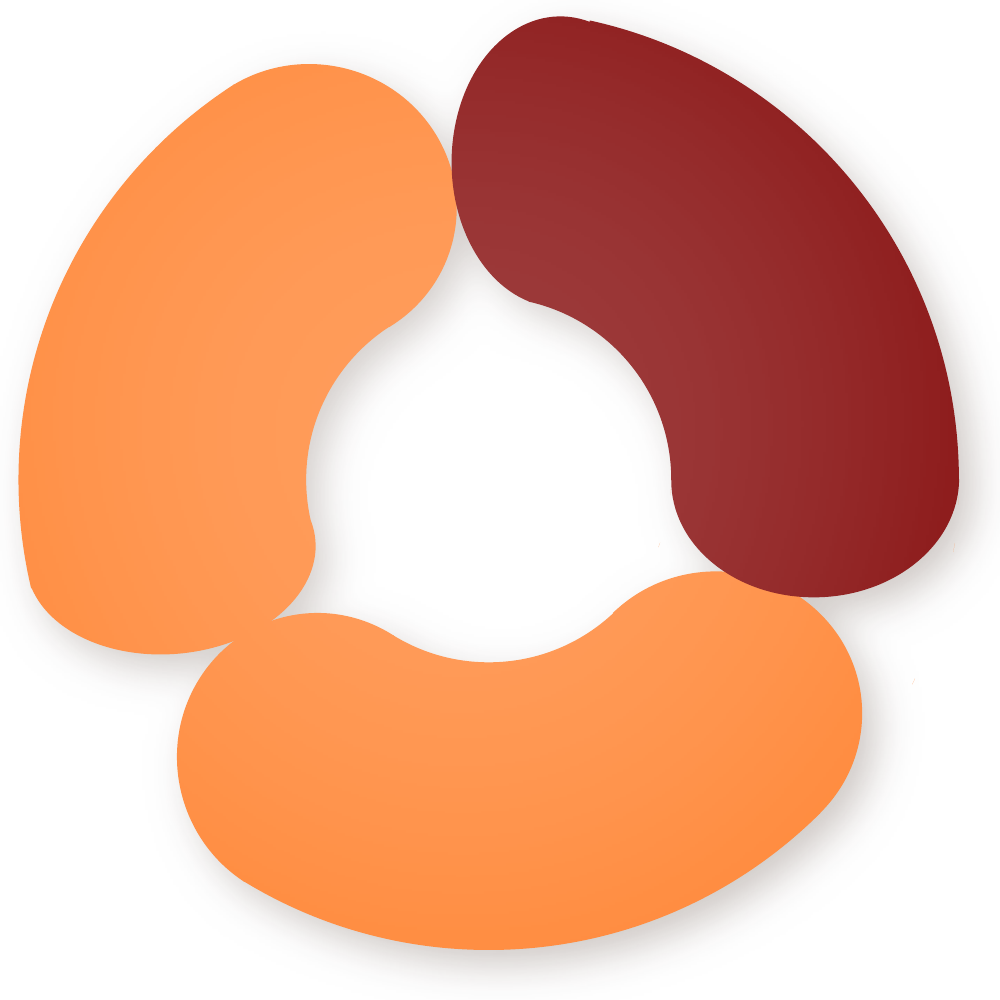}}
\put(0.5, 2.2){$\sigma_0$}
\put(3.2, 2){$\times$}
\put(3, 2.2){$\sigma$}
\put(1.5, 2.1){$\{s=0\}$}
\put(1.5, 1.2){$\{\Sigma=0\}$}
\put(5.8,1.6){\begin{turn}{70}$\{\Xi=0\}$\end{turn}}
\put(6.1,1.35){\begin{turn}{0}$\{t=0\}$\end{turn}}
\put(6.45,2.45){\begin{turn}{-60}$\{y_1=0\}$\end{turn}}
\put(6.7, 3){$\sigma$}
\put(7.6, 2){$\sigma_0$}
\put(7.9, 1.75){$\times$}
\end{picture}
 \caption{Schematic behavior of the fiber geometry
 over the two non-holomorphic loci. On the left, the locus $\{ e=f=0\}$
 is depicted. $\sigma_0$ wraps the entire fiber component, while $\sigma$
 cuts out a single point. On the right, the locus $\{f=g=0\}$ is shown,
 where $\sigma$ becomes non-holomorphic and $\sigma_0$ cuts out a point
 in the same fiber component. Fiber components wrapped by a section
 are colored dark red.}
 \label{fig:fiber_geometry}
\end{figure}

Let us study the behavior of the elliptic fiber at these points more
carefully. For $f=e=0$, the Calabi-Yau equation becomes
\begin{align}
  s(gw^2 + wty_1 P'(sy_1, y_2) + t^2y_1^2Q''(sy_1, y_2))\equiv s\Sigma
  =0
\end{align}
where $P'=P/(sy_1)$, and $Q''=Q'/(sy_1)$, which are homogeneous
polynomials when $f$ and $e$ vanish, of degrees 1 and 2 respectively
in the $y_i$. We see that at this locus the elliptic fiber degenerates
into two components, given by $s=0$ and $\Sigma=0$. When $s=0$ we can
gauge fix $\bC_1^*$ and $\bC^*_2$ in~\eqref{eq:resolved-GLSM} by
setting $t=y_2=0$, so we end up with the $y_1,w$ coordinates, with
relative SRI $\langle wy_1\rangle$, and identified by the $\bC^*$
action $(y_1,w)=(\lambda y_1, \lambda w)$. This is the usual
description of $\bP^1$, as one could have expected from the fact that
$s=0$ was the blow-up divisor. The curve $\Sigma$ defines a degree
$(4,2,2)$ divisor on the ambient space, and a simple adjunction
computation gives then that $\Sigma$ has genus $0$, i.e. it is also a
$\bP^1$. More explicitly
\begin{align}
  \begin{split}
    \chi(\Sigma) & = \int_\Sigma c_1(T\Sigma) = \int_A (c_1(TA) -
    \Sigma)\Sigma \\
    & = - \int_A [0,0,1] \wedge [4,2,2] = 2 \int_A [w]\wedge [s]\\
    & = 2 
  \end{split}
\end{align}
where $A$ denotes the ambient toric space~\eqref{eq:resolved-GLSM},
and on the second line we have denoted the divisor classes by their
toric weights.

These two spheres intersect over a point: setting $s=0$ (and thus
$y_2=t=1$) in the equation for $\Sigma$ we get:
\begin{align}
  gw^2 + wy_1P'(0,y_2) + y_1^2 Q''(0,y_2)=0\, .
\end{align}
This is a quadratic on the exceptional $\bP^1$, which has exactly two
solutions. So we recover the usual picture of the $T^2$ fiber
degenerating into two spheres, touching at two points. The rational
section $\sigma_0$ wraps one of the two sphere components, namely $s=0$.

A similar analysis holds for $\sigma$. Setting $g=f=0$
on~\eqref{eq:resolved-quartic} the Calabi-Yau equation factorizes as
\begin{align}
  y_1t(wsP'(sy_1, y_2) + tQ'(sy_1, y_2)) \equiv y_1t\,\Xi = 0 \, .
\end{align}
We find that there are three components in the fiber. By the same
kind of analysis as above we find that they are $\bP^1$s: for $y_1=0$
and $t=0$ this is immediate by looking
to~\eqref{eq:resolved-GLSM}. One also has that $\Xi=0$ is an equation
of degree $(3,1,0)$, and an adjunction computation gives that it has
genus 0.

The intersections between the three spheres can be computed easily,
with the result that any two of the three spheres intersect at exactly
one point. Our section $\sigma$ wraps the $y_1=0$ component.
A summary of the fiber geometry is contained in figure \ref{fig:fiber_geometry}.

\subsection{Physics of the conifold transition} \label{ss:conifolds}

The low energy description of the conifold transition is well
understood, starting with the seminal paper by Strominger
\cite{Strominger:1995cz} (see also \cite{Greene:1995hu,Greene:1996cy},
and \cite{Mohaupt:2004pq} for a treatment specialized to M-theory on
Calabi-Yau threefolds), so we will be brief here.

The basic physics mechanism in effective field theory language is
simply a Coulomb/Higgs branch transition: at the conifold point there
are a number of massless hypermultiplets, coming from M2 branes
wrapped on the collapsed $S^2$ cycles. We can smooth the conifold
points in two ways: deformation or resolution. On the resolved side
the 2-spheres take finite size, and this corresponds to making the M2
states massive. In field theoretic terms, this mass terms are
associated with the introduction of (geometry dependent) mass terms
for the hypermultiplets. More in detail, in M-theory compactified on a
smooth Calabi-Yau threefold $\hX$, there are $n_H = h^{2,1}(\hX)+1$
hypermultiplets, and $h^{1,1}(\hX)$ $U(1)$ gauge fields. A particular
combination of these belongs to the gravity multiplet, and the other
$n_V = h^{1,1}(\hX)-1$ $U(1)$ fields belong to vector
multiplets. These vector multiplets have a real bosonic scalar
component. The size of the resolved 2-spheres (keeping the overall
size of the Calabi-Yau threefold fixed) is precisely encoded in the
values of these scalars, so resolving the conifold singularities
corresponds to going into a Coulomb branch of the field theory.

On the other hand, there is a Higgs branch obtained by giving VEVs to
the massless hypermultiplets. This corresponds to smoothing out the
conifold singularities by complex deformations. Since the massless
hypermultiplets are naturally charged under the $U(1)$ symmetries (M2
branes couple electrically to $C_3$), giving a VEV will make some of
the $U(1)$ vector multiplets massive.

There is a simple relation between the counting of massless fields in
the five-dimensional theory and the Hodge numbers of the spaces related by the
conifold transition. Assume that there are $P$ 2-spheres degenerating
at $P$ conifold points. Typically not all of these 2-spheres are
linearly independent, but there are $R$ homology relations between
them (so $P-R$ independent classes vanish). Writing down the low
energy effective field theory for the hypermultiplets at the conifold
point, one can easily see \cite{Greene:1995hu,Greene:1996cy} that
there are precisely $R$ flat directions of the hypermultiplets, along
which one can Higgs them. A generic such Higgsing will then give mass
to $P-R$ vectors. All in all, M-theory on the resolved Calabi-Yau
threefold $\hX$ gives rise to a massless spectrum with
$(n_H(\hX),n_V(\hX))=(h^{2,1}(\hX)+1, h^{1,1}(\hX)-1)$. At the
conifold point, $P$ extra hypers become massless: $(n_H^0, n_V^0) =
(h^{2,1}(\hX)+1+P, h^{1,1}(\hX)-1)$. Higgsing then removes $P-R$
hyper-vector pairs: $(n_H(\tX), n_V(\tX))=(h^{2,1}(\hX)+1+R,
h^{1,1}(\hX)-1-P+R)$. On the other hand, these numbers are just
$h^{2,1}(\tX)+1$ and $h^{1,1}(\tX)-1$, respectively, so we learn that
the conifold transition acts on the Hodge numbers as
\begin{align}
  (h^{2,1}(\tX), h^{1,1}(\tX)) = (h^{2,1}(\hX)+R, h^{1,1}(\hX)-P+R)\, .
\end{align}
This formula will provide a nice consistency check that we are
identifying the geometry properly in our forthcoming examples (in our
examples, $P-R=1$, so $h^{1,1}(\hX)-h^{1,1}(\tX)=1$). A simple
quantity to check, in particular, is the difference in Euler numbers
\begin{align}
  \label{eq:conifold-count}
  \begin{split}
    \chi(\hX)-\chi(\tX) & = 2(h^{2,1}(\tX) - h^{2,1}(\hX)) - 2(h^{1,1}(\tX) - h^{1,1}(\hX))\\
    & = 2P
  \end{split}
\end{align}
giving the number of conifold points involved in the transition.

\subsection{Explicit examples with base $\mathbb{P}^2$} \label{ss:pairs_base_p2}

Having described the general setup for our main class of examples, we
are now ready to construct a number of examples of conifold
transitions removing the section. For simplicity, we will stay with a
$\bP^2$ base.

Let us start on the deformed side $\tX$. The set of Calabi-Yau
threefolds $T^2$-fibered over $\bP^2$ can be described as
hypersurfaces on the toric ambient space described by the GLSM
\begin{align}
  \label{eq:P112-fibration}
  \begin{array}{c|ccccccc}
    & x_1 & x_2 & x_3 & y_1 & y_2 & w & t\\
    \hline
    \bC^*_1 & 1 & 1 & 1 & 0 & a & b & 0\\
    \bC^*_2 & 0 & 0 & 0 & 1 & 1 & 2 & 0\\
    \bC^*_3 & 0 & 0 & 0 & 0 & 0 & 1 & 1
  \end{array}
\end{align}
The last four coordinates parametrize the fiber $\wh{\bP^{1,1,2}}$,
while the first three coordinates parametrize the base $\bP^2$. The
fibration map $\pi\colon X\to\bP^2$ simply ``forgets'' about the last
four coordinates of any point in $X$. In principle the last four
entries in the first row (the charges of $y_1,y_2,w,t$ under
$\bC^*_1$) can be arbitrary integers, but it is easy to convince
oneself that by redefining (if necessary) the $y_i$ and the $\bC^*_i$,
any such fibration can be brought to the canonical
form~\eqref{eq:P112-fibration}, with $a\geq 0$.

The generic equation in these variables is given
by~\eqref{eq:quartic}. In order to have a Calabi-Yau
threefold,~\eqref{eq:quartic} must be a homogeneous polynomial of
degree $(3+a+b, 4, 2)$. Tracing the definitions above, this implies
that the interesting coefficients of~\eqref{eq:quartic} are
homogeneous functions on the $x_i$ of degrees
\begin{align}
  \deg(a) & = 3-3a+b\\
  \deg(e) & = 3-2a+b\\
  \deg(f) & = 3-a\\
  \deg(g) & = 3+a-b\, .
\end{align}

There are a finite number of allowed values for $(a,b)$, obtained by
imposing that all the coefficients of~\eqref{eq:quartic} be
holomorphic functions on the $x_i$ (in particular, there should be no
poles). These conditions define a polygon in the $(a,b)$ plane, as
pointed out in \cite{Braun:2013nqa, Cvetic:2013uta}, and the different
cases, given in table~\ref{table:deformed-CYs}, correspond to integral
points of this auxiliary polygon.
\begin{table}
  \begin{align*}
    \begin{array}{c|cccccc}
      (a,b) & h^{1,1}(\tX) & h^{2,1}(\tX) & \deg(a) & \deg(e) & \deg(f) &
      \deg(g)\\
      \hline
      \hline
      (0,3) & 2 & 128 & 6 & 6 & 3 & 0\\
      (1,4) & 2 & 132 & 4 & 5 & 2 & 0\\
      (2,5) & 2 & 144 & 2 & 4 & 1 & 0\\
      (0,-2) & 3 & 59 & 1 & 1 & 3 & 5\\
      (0, -1) & 3 & 65 & 2 & 2 & 3 & 4\\
      (0,0) & 3 & 75 & 3 & 3 & 3 & 3\\
      (0,1) & 3 & 89 & 4 & 4 & 3 & 2\\
      (0,2) & 3 & 107 & 5 & 5 & 3 & 1\\
      (1,0) & 3 & 69 & 0 & 1 & 2 & 4\\
      (1,1) & 3 & 79 & 1 & 2 & 2 & 3\\
      (1,2) & 3 & 93 & 2 & 3 & 2 & 2\\
      (1,3) & 3 & 111 & 3 & 4 & 2 & 1\\
      (2,3) & 3 & 105 & 0 & 2 & 1 & 2\\
      (2,4) & 3 & 123 & 1 & 3 & 1 & 1\\
      (3,6) & 3 & 165 & 0 & 3 & 0 & 0\\
      (0, -3) & 6 & 60 & 0 & 0 & 3 & 6
    \end{array}
  \end{align*}

  \caption{Hodge numbers and polynomials degrees for various
    fibrations over $\bP^2$}

  \label{table:deformed-CYs}
\end{table}

There are some interesting features in this table. Notice that the
first three entries have $\deg(g)=0$. Taking $g$ a generic non-zero
constant, we find that $Q$ becomes a holomorphic section, since the
$f=g=0$ locus does not exist anymore. Similarly, for the $(0, -3)$
example the $\sigma_0$ section is holomorphic, and for the $(3,6)$ example
both sections are holomorphic. In the rest of the cases both sections
are rational.

The resolved side $\hX$ is given by hypersurfaces on toric ambient
spaces described by GLSMs of the form

\begin{align}
  \label{eq:resolved-P112-fibration}
  \begin{array}{c|cccccccc}
    & x_1 & x_2 & x_3 & y_1 & y_2 & w & t & s\\
    \hline
    \bC^*_1 & 1 & 1 & 1 & 0 & a & b & 0 & 0\\
    \bC^*_2 & 0 & 0 & 0 & 1 & 1 & 2 & 0 & 0\\
    \bC^*_3 & 0 & 0 & 0 & 0 & 0 & 1 & 1 & 0\\
    \bC^*_4 & 0 & 0 & 0 & 1 & 0 & 1 & 0 & -1
  \end{array}
\end{align}
As before, in principle we could have given a charge to $s$ under
$\bC_1^*$, but there is always a way of redefining the fields and
$\bC^*$ symmetries in order to set this charge to 0. Imposing that the
coefficients of~\eqref{eq:resolved-quartic} are sections of line
bundles of non-negative degree on the $\bP^2$ base, one finds 31
different possible values for $(a,b)$. All those in
table~\ref{table:deformed-CYs} are included, and in addition there are
a few models which are only possible on the resolved side, since the
blow-up fixes the coefficient of the $y_2^4$ term in $Q$ to vanish, so
there is one less constraint. We will only be interested in the ones
coming from conifold transitions on $\tX$.

Identifying the models in the canonical way, we can immediately
compute the Hodge numbers of the resolved spaces using PALP, for
instance, the results are given in
table~\ref{table:resolved-CYs}. Computing from here the expected
number of conifold points, with the results shown in the last column
of table~\ref{table:resolved-CYs}, one sees easily by comparing with
the values in table~\ref{table:deformed-CYs} that in all cases the
expected number of conifold points precisely agrees with the
expectation from the discussion given above:
\begin{align}
  \frac{1}{2}(\chi(\hX)-\chi(\tX)) = \deg(e)\cdot \deg(f)\, .
\end{align}

\begin{table}
  \begin{align*}
  \begin{array}{c|ccc|ccccc}
    (a,b) & h^{1,1}(\hX) & h^{2,1}(\hX) & P & H(\rep{1}_2) & H(\rep{1}_4) & H(\rep{2}_1) & H(\rep{2}_3) & H(\rep{3}_0)\\
    \hline
    \hline
    (0,3) & 3 & 111 & 18 & 144 & 18 & 0 & 0 & 0 \\
    (1,4) & 3 & 123 & 10 & 140 & 10 & 0 & 0 & 0 \\
    (2,5) & 3 & 141 & 4 & 128 & 4 & 0 & 0 & 0 \\
    (0,-2) & 4 & 57 & 3 & 64 & 3 & 55 & 15 & 6 \\
    (0, -1) & 4 & 60 & 6 & 76 & 6 & 52 & 12 & 3 \\
    (0,0) & 4 & 67 & 9 & 90 & 9 & 45 & 9 & 1 \\
    (0,1) & 4 & 78 & 12 & 106 & 12 & 34 & 6 & 0 \\
    (0,2) & 4 & 93 & 15 & 124 & 15 & 19 & 3 & 0 \\
    (1,0) & 4 & 68 & 2 & 72 & 2 & 56 & 8 & 3 \\
    (1,1) & 4 & 76 & 4 & 86 & 4 & 48 & 6 & 1 \\
    (1,2) & 4 & 88 & 6 & 102 & 6 & 36 & 4 & 0 \\
    (1,3) & 4 & 104 & 8 & 120 & 8 & 20 & 2 & 0 \\
    (2,3) & 4 & 104 & 2 & 90 & 2 & 38 & 2 & 0 \\
    (2,4) & 4 & 121 & 3 & 108 & 3 & 21 & 1 & 0 \\
    (3,6) & 3 & 165 & 0 & 108 & 0 & 0 & 0 & 0 \\
    (0, -3) & 6 & 60 & 0 & - & - & - & - & -
  \end{array}
  \end{align*}

  \caption{Hodge numbers and chiral spectra for the resolved versions
    of the manifolds in table~\ref{table:deformed-CYs}. All $U(1)$
    charges have been rescaled by $2$. $P$ denotes the expected number
    of conifold points, obtained from~\eqref{eq:conifold-count}. The
    last entry in the table corresponds to a space with many
    non-torically realized divisors, so we will not analyze it here.}

  \label{table:resolved-CYs}
\end{table}

In table~\ref{table:resolved-CYs} we summarize information about the
models obtained by resolving the manifolds from
table~\ref{table:deformed-CYs}, including the chiral spectrum in six
dimensions, obtained via the techniques described in
\cite{Grimm:2011fx, Grimm:2013oga}. Here
$H(\cR)$ denotes the net amount of chiral matter (six-dimensional hypers) in the
representation $\cR$. We denote the representation by ${\bf N}_m$, where
${\bf N}$ is the representation under the gauge group $SU(2)$ (to be
explained below), and $m$ the $U(1)$ charge. We define the divisor
class generating the $U(1)$ charge by~\cite{Grimm:2013oga}
\begin{align}
  \label{eq:DU(1)}
  D_{U(1)} = 2\sigma - 2\sigma_0 - 4\pi^*c_1(TB) + E\, .
\end{align}
We have denoted $\pi\colon \hX\to\bP^2$ the fibration map, $\pi^*$ its
pullback to cohomology on $X$, $\sigma,\sigma_0$ denote the extra
section and the zero section described above, and $E$ is the divisor
associated with the Cartan of $SU(2)$. The single manifold with
$h^{1,1}(\hX) = 6$ has three divisors that do not descend from the
ambient space and it is unclear what the full gauge group and matter
spectrum are, so we will not analyze it here. Lastly, let us remark
that we find that
\begin{align}
  H(\rep{1}_4) = \frac{1}{2}(\chi(\hX)-\chi(\tX)) = [e] \cdot [f]
\end{align}
which strongly suggests that it is precisely the $\rep{1}_4$
multiplets that are involved in the conifold transition.

The existence of an $SU(2)$ symmetry in the cases with
$h^{1,1}(\hX)>3$ can be argued for as follows. Consider the $g=0$
locus on the base (this is only possible if $\deg(g)>0$). Over this
divisor, the Calabi-Yau equation becomes
\begin{align}
  \wt{\phi}|_{g=0} = t(wP + ty_1Q') \equiv t\Lambda = 0\, .
\end{align}
We see that over this divisor on the base the $T^2$ factorizes. The
$t=0$ piece defines a $\bP^1$, and it is not hard to prove that
$\Lambda=0$ is also a $\bP^1$, intersecting $t=0$ at two points. This
is the familiar affine $SU(2)$ structure over a zero of the
discriminant, so we expect a $SU(2)$ enhancement over $g=0$. A short
computation shows, in addition, that the section $\sigma_0$ intersects
$\Lambda$ at a point, and $\sigma$ intersects $t=0$ at a point. Since
we chose $\sigma_0$ as our zero section, we interpret the component
not intersecting it, namely $t=0$, as the one associated with the $W$
bosons enhancing the gauge symmetry to $SU(2)$. All in all, we learn
that $E$ in~\eqref{eq:DU(1)} is just $\{t=0\}\cap\{\wt{\phi}=0\}$, or
$[t]$ in brief (abusing notation slightly).

We are in fact in a position to compute the charges of some of the
multiplets in table~\ref{table:resolved-CYs} from first principles. We
start by discussing the $\rep{1}_4$ multiplets, which are the main
actors in the conifold transition. The other representations can be
obtained analogously, with some extra effort. Since these
representations are less directly relevant for the conifold
transition, we demote their discussion to appendix~\ref{app:matter}.

We claim that the $\rep{1}_4$ multiplets comes from $f=e=0$. We have
explained above that when $f=e=0$ the fiber becomes split into two
components, given by $\{s=0\}\cup\{\Sigma=0\}$. Since $st$ belongs in
the Stanley-Reisner ideal, the hyper wrapping $s=0$ has no charge
under the $SU(2)$ symmetry. Its charge under the $U(1)$ is given by
\begin{align}
  \label{eq:Q-Cs}
  Q_{U(1)} = \cC_s \cdot (2\sigma - 2\sigma_0 - 12[x_1] + [t])\, .
\end{align}
We have denoted by $\cC_s$ the component of the fiber over $f=e=0$
given by $s=0$, and we used the fact that $[x_1]$ is the pullback of
the hyperplane on $\bP^2$. Since $x_1=0$ will generically not
intersect $f=e=0$, we have $\cC_s\cdot [x_1]=0$. Similarly, since $st$
is in the Stanley-Reisner ideal, $\cC_s\cdot [t]=0$. We already
determined above that $\sigma$ intersects $\cC_s$ at a point, so
$\cC_s\cdot \sigma = 1$. On the other hand, $\sigma_0$ becomes rational at
$f=e=0$, so the calculation is less straightforward. Consider the
total class of the (factorized) $T^2$ fiber, given by $\cC_s +
\cC_\Sigma$, with the last component being the $\Sigma=0$ locus. Since
the total fiber can move as a holomorphic divisor into a smooth $T^2$,
which intersects $\sigma_0$ at a point, it must be the case that $(\cC_s
+\cC_\Sigma)\cdot \sigma_0 = 1$. On the other hand, on the factorized locus
it is clear that $\cC_\Sigma\cdot \sigma_0 = 2$ (the two points where the
$\bP^1$ components touch). So we conclude $\cC_s\cdot \sigma_0 =
-1$. Substituting all this into~\eqref{eq:Q-Cs} we obtain
$Q_{U(1)}=4$, as claimed.

\subsection{Chern-Simons terms} \label{ss:cs_terms}
In this final subsection, we confirm geometrically that the Chern-Simons terms
of the theory obtained by compactifying M-theory on $\tX$ are in fact
related to the Chern-Simons terms of M-theory on $\hX$ as
described in equation \eqref{tk-def}. Instead of delving into concrete examples right away
and showing explicitly that this prescription is correct on a case by case basis,
let us make a general geometric argument first. As the Chern-Simons terms of the
five-dimensional models are given in terms of intersection numbers, we need to understand
how the intersection form on $\tX$ is obtained from the intersection form of $\hX$.
Fortunately for us, this was studied long ago, see for example \cite{Mohaupt:2004pq}. Denoting by $\cK_i$,
 $i=1,\dots, h^{1,1}(\hX)$ a basis of the K\"ahler cone on $\hX$ and by $\widetilde{\cK}_i$,
$i=1,\dots,h^{1,1}(\tX)$ the corresponding K\"ahler cone basis on $\tX$,
we choose the $\cK_i$
such that under the conifold transition they are mapped to divisors on $\tX$ according to
\begin{align} \label{e:conifold_map}
\cK_i \mapsto \begin{cases}
 \widetilde{\cK}_i & \textrm{if } i \leq h^{1,1}(\tX) \\
 0 & \textrm{otherwise.} \end{cases}
\end{align}
Then the intersection numbers of the $\widetilde{\cK}_i$ on $\tX$ are the same as of
the $\cK_i$ on $\hX$, i.e.
\begin{equation} \label{e:conifold_map_intersections}
\widetilde{\cK}_i \cdot \widetilde{\cK}_j \cdot \widetilde{\cK}_k = \cK_i \cdot \cK_j \cdot \cK_k\,.
\end{equation}
Put differently, the intersection form on $\tX$ is obtained by restricting the intersection form
on $\hX$. That is, given expressions for the volumes $\mathcal{V}$
 and $\widetilde{\mathcal{V}}$ of $\hX$ and $\tX$
in terms of the K\"ahler parameters $v^i$ and $\tilde{v}^i$, one has that
\begin{align} \label{e:conifold_map_v}
\widetilde{\mathcal{V}} = \mathcal{V} (v^1 = \tilde{v}^1, \dots, 
v^{h^{1,1}(\tX)}=\tilde{v}^{h^{1,1}(\tX)}, 0, \dots)\,.
\end{align}
Presented with this simple relation between triple intersections on $\hX$ and $\tX$, let
us now return to the discussion of the Chern-Simons terms of M-theory on $\tX$.
Given two independent sections on $\hX$ we know that only a certain linear combination
$D_{U(1)}$ is left untouched by the conifold transition -- the other $U(1)$-divisor
is eliminated as the corresponding gauge field gains a mass term. Identifying the surviving
$U(1)$ amounts to making the same clever choice of basis as for the $\cK_i$ above. Then,
equation \eqref{e:conifold_map_intersections} tells us that the intersection numbers of the
surviving $U(1)$-divisor are \emph{precisely the same} as on the resolved side.
Therefore, we are left with two questions to examine in our specific examples, namely:
\begin{enumerate}
\item Which divisor $D_{U(1)}$ survives the conifold transition?
\item Why is $D_{U(1)} \cdot c_2(\hX) = \tilde{D}_{U(1)} \cdot c_2(\tX)$?
\end{enumerate}
In subsection \ref{sec:eff_circle-flux} we gave a general argument for how to
identify $D_{U(1)}$ and, in fact, we will show explicitly that this prescription
does in fact select the correct divisor for the examples below. The second
point is more difficult to answer generally, but we can confirm it on a case
by case basis.

Put in a nutshell, we have explained generally that after a clever change of basis
the Chern-Simons terms of the theories corresponding to $\hX$ and $\tX$
are simply obtained by "dropping" the massive $U(1)$. Of course,
one can also confirm this statement explicitly through the calculation
of intersection numbers and in the remainder of this section we will
perform an example calculation.

\subsubsection{A close look at the model with $(a, b) = (0, 3)$}
For concreteness, let us study the manifold with $(a,b) = (0,3)$, beginning
on the resolved side. We find that the Mori cone is generated by the three curves
  \begin{align}
  \begin{array}{c||ccccccc}
     & x_1 & x_2 & x_3 & y_1 & y_2 & w & s \\
    \hline
    \hline
    \cC_1 & 1 & 1 & 1 & -3 & 0 & 0 & 3\\
    \cC_2 & 0 & 0 & 0 & -1 & 1 & 0 & 2 \\
    \cC_3 & 0 & 0 & 0 & 1 & 0 & 1 & -1\\
  \end{array}
  \end{align}
and we can hence choose
\begin{align}
 \cK_1 = x_1\,, \quad \cK_2  = y_2\,, \quad \cK_3 = w
\end{align}
as a basis of the K\"ahler cone satisfying $\cK_i \cdot \cC^j = \delta_i^j$. Expressing the K\"ahler form $J = \sum_{i=1}^3 v^i [\cK_i]$
in terms of two-forms dual to these divisors, one finds that the overall volume of the Calabi-Yau can be written as
\begin{align}
 \cV = (v^{1})^{2} v^{2} + \frac{3}{2} (v{1})^{2} v^{3} + 6 v^{1} v^{2} v^{3} +
\frac{15}{2} v^{1} (v^{3})^{2} + 9 v^{2} (v^{3})^{2} + \frac{21}{2}
(v^{3})^{3}\,.
\end{align}
Let us turn to the two divisors generating the $U(1)$ symmetries in five dimensions. One is obtained by
appropriately shifting the zero section \cite{Morrison:2012ei, Grimm:2013oga}, while the other
can be computed by applying the Shioda map to the other section. Naturally, a different choice
of zero section will lead to interchanged results for the divisor expansions. Since the
resulting physics remain unaffected, we choose the divisor $s=0$, or $\sigma_0$ in the notation
of subsection \ref{ss:pairs_base_independent}, as the zero section during the rest of this discussion.
Note that in this particular basis the divisors generating the two $U(1)$s have the expansion
\begin{align}
  D_{0} =  \frac{9}{2} \cK_1 + 2 \cK_2 - \cK_3\,,\quad D_{1} = -24 \cK_1 - 6 \cK_2 + 4 \cK_3\,.
\end{align}
Now we discuss the deformed manifold $\tX$. Its Mori cone is spanned by
\begin{align} \label{e:Mori_cone_0_3}
  \begin{array}{c||ccccccc}
     & x_1 & x_2 & x_3 & y_1 & y_2 & w \\
    \hline
    \hline
    \tilde{\cC}_1 & 1 & 1 & 1 & 0 & 0 & 3\\
    \tilde{\cC}_2 & 0 & 0 & 0 & 1 & 1 & 2 \\
  \end{array}
  \end{align}
and a good choice of K\"ahler basis is for example given by
\begin{align}
 \tilde{\cK}_1 = x_1\,, \quad \tilde{\cK}_2 = y_2\,.
\end{align}
Then the volume of the deformed manifold is
\begin{align}
 \tilde{\cV} = (\tilde{v}^1)^2 \tilde{v}^2\,.
\end{align}
Obviously, the intersection rings of $\hX$ and $\tX$ are related as in equation \eqref{e:conifold_map_v}, with
$\cK_3$ the divisor eliminated during the conifold transition. Up to an overall rescaling, there is hence
a unique combination of $D_0$ and $D_1$ that is left invariant under the conifold map, namely
the one not containing $\cK_3$. It is \footnote{Note that in subsection \ref{sec:eff_circle-flux} we denoted the
$U(1)$-divisor remaining massless by $\widetilde{D}_0$. Here we call it $D_{U(1)}$ to emphasize that it not
necessarily a divisor on $\tX$.}
\begin{align}
D_{U(1)} \sim 4 D_{0} + D_{1}\,.
\end{align}
Since we rescaled the six-dimensional $U(1)$ divisor on $\hX$ by $\lambda=2$, this is precisely the expression
that we expect from equation \eqref{eq:surviving_divisor}. Lastly, we can check by
explicit computation that $D_{U(1)} \cdot c_2(\hX) = \widetilde{D}_{U(1)} \cdot c_2(\tX)$.

\subsubsection{A close look at the model with $(a,b) = (0, -2)$}
As a second example, we repeat the analysis for one of the models that contain
an additional $SU(2)$ factor to show that the above discussion is independent of the
existence of additional gauge group factors. Again, we begin with the resolved manifold $\hX$,
whose Mori cone is this time spanned by the curves
  \begin{align}
  \begin{array}{c||cccccccc}
     & x_1 & x_2 & x_3 & y_1 & y_2 & w & s & t \\
    \hline
    \hline
    \cC_1 & 1 & 1 & 1 & 0 & 0 & -2 & 0 & 0\\
    \cC_2 & 0 & 0 & 0 & 1 & 1 & 0 & 0 & -2 \\
    \cC_3 & 0 & 0 & 0 & 1 & 0 &1 & -1 & 0 \\
    \cC_4 & 0 & 0 & 0 & -1 & 0 & 0 & 1 & 1 \\
  \end{array}
  \end{align}
and we pick
\begin{align}
 \cK_1 = x_1\,, \quad \cK_2 = y_2\,, \quad \cK_3 = t + 2 y_2\,, \quad \cK_4 = w + 2 x_1
\end{align}
as the basis of the K\"ahler cone. The volume of the resolved manifold is then
\begin{align}
 \cV &= (v^{1})^{2} v^{2} + 2 (v^{1})^{2} v^{3} + 5 v^{1} v^{2} v^{3} + 5 v^{1}
(v^{3})^{2} + 5 v^{2} (v^{3})^{2} + \frac{10}{3} (v^{3})^{3} + \frac{3}{2}
(v^{1})^{2} v^{4} \nonumber \\
& \quad + 5 v^{1} v^{2} v^{4} + 10 v^{1} v^{3} v^{4} + 10 v^{2}
v^{3} v^{4} + 10 (v^{3})^{2} v^{4} + \frac{7}{2} v^{1} (v^{4})^{2} \nonumber \\
& \quad + 5 v^{2} (v^{4})^{2} + 10 v^{3} (v^{4})^{2} + \frac{7}{3} (v^{4})^{3}\,.
\end{align}
Choosing $\sigma_0 = \{s=0\}$ as zero section and expanding the $U(1)$
divisors of the five-dimensional theory in a basis of $\cK_i$ one finds
\begin{align}
 D_{0} = \frac{3}{2} \cK_1 + \cK_3 - \cK_4\,, \quad D_1 = -12 \cK_1 -3 \cK_3 + 4 \cK_4\,.
\end{align}
Additionally, there is a third $U(1)$ which is enhanced to the non-Abelian $SU(2)$ factor
in the F-theory limit. We denote it by $E$ and its expansion reads
\begin{align} \label{e:su2_example_E}
 E =  -2 \cK_1 + \cK_2\,.
\end{align}
Changing to the deformed manifold $\tX$ corresponding to F-theory with a massive $U(1)$,
we find that its Mori cone is generated by
  \begin{align}
  \begin{array}{c||ccccccc}
     & x_1 & x_2 & x_3 & y_1 & y_2 & w & t \\
    \hline
    \hline
    \tilde{\cC}_1 & 1 & 1 & 1 & 0 & 0 & -2  & 0\\
    \tilde{\cC}_2 & 0 & 0 & 0 & 1 & 1 & 0  & -2 \\
    \tilde{\cC}_3 & 0 & 0 & 0 & 0 & 0 & 1 & 1 \\
  \end{array}
  \end{align}
and we parametrize the K\"ahler form in terms of two-forms Poincar\'e-dual to
\begin{align}
 \tilde{\cK}_1 = x_1\,, \quad \tilde{\cK}_2 = y_2 \,, \quad \tilde{\cK}_3 = t+2 y_2\,.
\end{align}
The volume of $\tX$ is given by
\begin{align}
 \widetilde{\cV} &= (\tilde{v}^{1})^{2} \tilde{v}^{2} + 2 (\tilde{v}^{1})^{2} \tilde{v}^{3}
 + 5 \tilde{v}^{1} \tilde{v}^{2} \tilde{v}^{3} + 5 \tilde{v}^{1} (\tilde{v}^{3})^{2}
 + 5 \tilde{v}^{2} (\tilde{v}^{3})^{2} + \frac{10}{3} (\tilde{v}^{3})^{3}
\end{align}
and one can see that it is obtained by restricting the volume of the resolved phase
according to
\begin{align}
 \tilde{\cV} = \cV \rvert_{v^4=0, v^i = \tilde{v}^i}\,.
\end{align}
Consequently, we see that the above choice of $\cK_i$ is again a good one in the sense of
equations \eqref{e:conifold_map} and \eqref{e:conifold_map_v} and one transitions
from $\hX$ to $\tX$ by dropping $\cK_4$. Since equation \eqref{e:su2_example_E}
does not contain $\cK_4$, we observe that it is left untouched by the conifold transition
and does not take part in the mixing involving the remaining two $U(1)$s. Requiring again
that the surviving $U(1)$ must not contain $\cK_4$, one finds that, up to an overall rescaling,
it is given by
\begin{align}
D_{U(1)} = 4 D_0 + D_1\,,
\end{align}
which, as before, matches the prescription of \eqref{eq:surviving_divisor} with $\lambda=2$.
In summary, we find that the discussion of the case with additional $SU(2)$ gauge
symmetry is almost identical to the one of the simpler models with only Abelian gauge groups.
As before, we identify a curve shrinking to zero volume in the conifold limit.
The intersection form of the deformed model is then obtained by dropping the divisor
dual to that curve from the intersection form of the resolved phase.
As the $SU(2)$ Cartan divisor does not contain the divisor that is eliminated
in the conifold transition, it does not mix with any of the other $U(1)$s during
the conifold transition.
Finally, one can again confirm that $D_{U(1)} \cdot c_2(\hX) = \widetilde{D}_{U(1)} \cdot c_2(\tX)$,
thereby showing that the Chern-Simons terms corresponding to the higher curvature terms
are matched as well.

\subsubsection{Explicit formulas for the Chern-Simons terms}
Technically, the previous discussion already ensures the matching of the Chern-Simons terms
as discussed in section \ref{sec:eff_circle-flux}. Nevertheless, it may be illuminating to
consider the discussion from a different angle. Let us therefore evaluate formulas
\eqref{1-loop_tildek000} and \eqref{1-loop_tildek}
for the examples at hand and show that they predict the correct intersection numbers.
Turning the discussion around, one can also use these relations to \emph{compute the spectrum
of $\tX$} without making use to the resolved manifold $\hX$.

\begin{table}[h]
  \begin{align*}
  \begin{array}{c|cccc|cc}
    (a,b) & V & H_{neutral} & H(\rep{1}_2) & H(\rep{1}_4) & \tilde{k}_{0} & \tilde{k}_{000} \\
    \hline
    \hline
    (0,3) & 1 & 112 & 144 & 18 & -168 & 432\\
    (1,4) & 1 & 124 & 140 & 10 &-128  & 304 \\
    (2,5) & 1 & 142 & 128 & 4 &-80 & 208\\
    (3,6) & 1 & 166 & 108 & 0 & -24 &  144\\
  \end{array}
  \end{align*}
  
  \caption{Spectra and Chern-Simons coefficients of $\widetilde{A}^0$ for the models with two sections and $h^{1,1} = 3$. Here,
  the Chern-Simons terms are obtained from the geometry and can be shown to match the field theory computation. All $U(1)$
    charges have been rescaled by $2$.}
    
  \label{table:cs_coefficients_h11_3}
\end{table}
This time, we restrict ourselves to models with purely Abelian gauge group, where
we know the spectrum to consist of $\rep{1}_2$ and $\rep{1}_4$ states.
Assuming furthermore that
\begin{align}
l_{\rep{1}_2} = 0\,, \qquad l_{\rep{1}_4} = 1
\end{align}
as is the case when $\hX$ has a non-holomorphic zero section (corresponding to $\sigma_0$
as above), the formulas for $\tilde{k}_{000}$ and $\tilde{k}_{0}$ simplify to
\begin{align}
\tk_{000} &= \frac{m^3}{120} \left(H - V - T - 3 \right) \nonumber \\
& \quad + \frac{1}{4} H(\rep{1}_2) \left( -4 n^2 m + 16 n^3 \sign(\rep{1}_2) \right) \nonumber \\
& \quad + \frac{1}{4} H(\rep{1}_4) \left( -4 m^3  - 208 n^2 m  +(384 n^3 + 48 n m^2 ) \sign(\rep{1}_4) \right)\,.
\end{align}
and
\begin{align}
\tk_0 &= \frac{m}{6} \left( H - V + 5 T + 15\right) \nonumber \\
& \quad + H(\rep{1}_2) (- 2 n \sign(\rep{1}_2)) 
+ H(\rep{1}_4) \left( 2 m - 12 n \sign(\rep{1}_4) \right)\,.
\end{align}
To be as concrete as possible, we plug in $n = -1$ and $m = 4$ as we found above and
use that for these manifolds $\sign(\rep{1}_2) = \sign(\rep{1}_4) = -1$ and $T=0$
to find
\begin{align}
\tk_{000} &= \frac{8}{15} \left( H - V - 3 \right) + 16 H(\rep{1}_4) \\
\tk_0 &= \frac{2}{3} ( H - V + 15) - 2 H(\rep{1}_2) - 4 H(\rep{1}_4)\,.
\end{align}
Evaluating the formulas, one easily confirms that they indeed match the intersection numbers
given in table \ref{table:cs_coefficients_h11_3}. Note that table \ref{table:cs_coefficients_h11_3}
contains the spectra of the F-theory models on the resolved manifolds $\hX$. However,
they can easily be translated to the case of a massive $U(1)$ corresponding to F-theory on $\tX$. F-theory on $\tX$ has
$H_{neutral}-1$ neutral hypermultiplets and $V=0$ massless vectors as shown in figure \ref{fig:big_picture}.
In six dimensions, the charged spectrum is the same on $\hX$ and $\tX$ with the difference
that the $U(1)$ field in F-theory on $\tX$ is massive. However, upon doing
the fluxed circle reduction to five dimensions, the $\rep{1}_4$ states with KK-level $\hat{n}=-1$ are neutral under
the mixed massless $U(1)$ gauge field $\widetilde{A}^0$ and must therefore be counted as additional
neutral states not counted by $h^{2,1}(\tX)$.

We remark that  these are the same results as one would get by starting
with the conjectured six-dimensional F-theory set-up with a massive $U(1)$. In fact,
by computing the Mori cones of $\hX$ and $\wt{X}$ one can show that the $\sign$ functions
for the states $\rep{1}_2$ and $\rep{1}_4$ agree in the deformed and the resolved
phases.

Finally, let us comment on directly computing spectra of F-theory models $\tX$ without section. In 
the examples studied, we gained an computational advantage by finding models $\hX$ with section that are related
to $\tX$ by conifold transitions. Ideally, however, one would like to compute the spectra of F-theory on
$\tX$ without making this detour. In general, this is going to be more difficult due to the fact that
there are less divisors on $\tX$ and therefore less intersection numbers to extract information from
even though the spectra are equally complicated. As it turns out, for cases with a single $U(1)$ there
are generally more unknown variables than equations obtained from matching the Chern-Simons terms.
However, if one also requires all anomalies to be canceled, it \emph{is possible} to compute the spectra
directly from $\tX$ for the cases presented here. Incorporating these methods into a general
approach by extending the variety of models studied here seems to be a promising
direction of study.

\section{Conclusions}

In this paper we studied the effective physics of F-theory compactifications on elliptically
fibered Calabi-Yau threefolds that do not have sections, but instead admit a bi-section.
Applying M- to F-theory duality, we found that an ordinary circle reduction of
a six-dimensional theory without $U(1)$ gauge symmetries is not sufficient to match M-theory
compactified to five dimensions. Instead, we claimed that there exists a six-dimensional $U(1)$ symmetry
made massive by a geometric St\"uckelberg mechanism.
Nevertheless, a simple circle reduction of this putative effective theory in six dimensions
is not yet sufficient to achieve a match with M-theory either. Due to the absence of a section,
non-trivial NS-NS and R-R fluxes appear along the circle direction used in the compactification
to five-dimensional and an axionic degree of freedom is shift-gauged by the respective Kaluza-Klein vector.
Caused by this additional gauging, the Kaluza-Klein vector and the massive $U(1)$ vector mix
in the fluxed circle compactification and a linear combination of both vectors remains massless
in the five-dimensional effective theory. Geometrically, this massless $U(1)$ vector is identified
with the bi-section of the genus-one fibration.

Having treated such set-ups generally, we presented a class of example geometries in section \ref{sec:examples}.
In this class of examples, we found that one can employ a conifold transition to pass from an
F-theory model without section to one that does admit a section. Geometrically, the manifolds with
bi-section correspond to deformations of the singular conifold points, while one obtains
the manifolds with sections by resolving the conifold singularities. As expected from the vast
physics literature on this subject, we confirm that physically, a certain set of states becomes
massless during the conifold transition and Higgs one of the two massless five-dimensional $U(1)$ gauge fields.
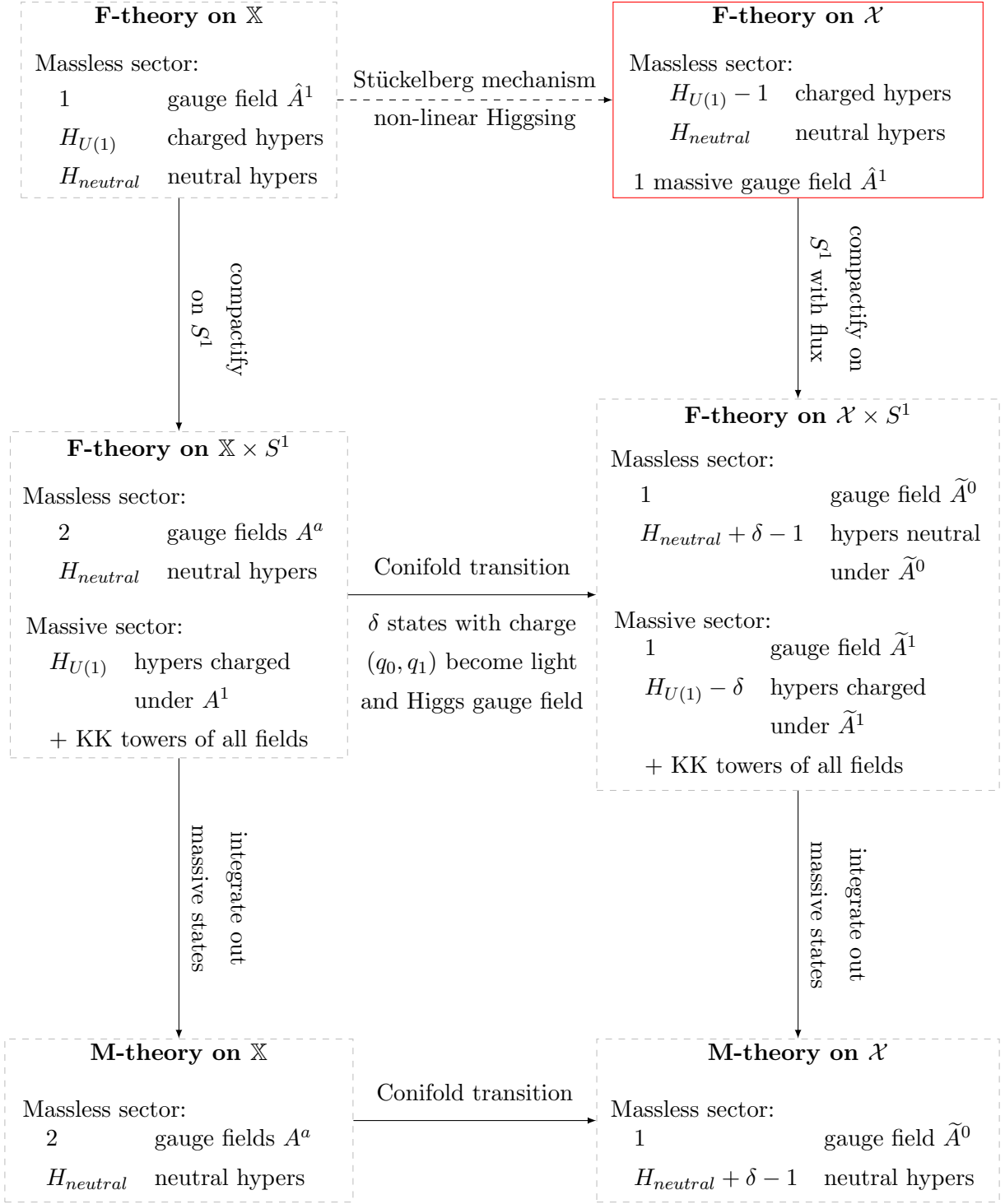
\begin{figure}[h]
\centering
  \begin{tikzpicture}
    
    \kbbox[90, 145,north,F_resolved,(0, 0)] {\textbf{F-theory on $\hX$}};
    \node [] () at (-1, 0.6) {Massless sector:};
    \node [] () at (0.2, -0.6) {\begin{tabular}{ll}
     1 & gauge field $\hat{A}^1$\\
     $H_{U(1)}$ & charged hypers \\
     $H_{neutral}$ & neutral hypers \\
  \end{tabular}};
  
  \kbboxred[90, 170,north,F_deformed,(10, 0)] {\textbf{F-theory on $\tX$}};
    \node [] () at (8.6, 0.6) {Massless sector:};
    \node [] () at (10.2, -0.2) {\begin{tabular}{ll}
     $H_{U(1)}-1$ & charged hypers \\
     $H_{neutral}$ & neutral hypers \\
  \end{tabular}};
    \node [] () at (9.4, -1.3) {$1$ massive gauge field $\hat{A}^1$};
    
  \draw[dashed, -latex] (F_resolved.east) -> 
  node[above]{St\"uckelberg mechanism} node[below]{non-linear Higgsing} (F_deformed.west);
  
  \kbbox[150, 155,north,F_resolved_S1,(0, -8)] {\textbf{F-theory on $\hX \times S^1$}};
    \node [] () at (-1.2, -6.4) {Massless sector:};
    \node [] () at (0.2, -7.3) {\begin{tabular}{ll}
     2 & gauge fields $A^a$\\
     $H_{neutral}$ & neutral hypers \\
  \end{tabular}};
    \node [] () at (-1.2, -8.5) {Massive sector:};
    \node [] () at (0.0, -9.7) {\begin{tabular}{ll}
     $H_{U(1)}$ & hypers charged\\
     & under $A^1$\\
     \multicolumn{2}{l}{+ KK towers of all fields}
  \end{tabular}};
  
  \kbbox[180, 185,north,F_deformed_S1,(10, -8)] {\textbf{F-theory on $\tX \times S^1$}};
    \node [] () at (8.3, -5.8) {Massless sector:};
    \node [] () at (10.2, -7) {\begin{tabular}{ll}
     1 & gauge field $\widetilde{A}^0$\\
     $H_{neutral} +\delta -1$ & hypers neutral \\
     & under $\widetilde{A}^0$
  \end{tabular}};
    \node [] () at (8.3, -8.4) {Massive sector:};
    \node [] () at (9.8, -9.8) {\begin{tabular}{ll}
     $1$ & gauge field $\widetilde{A}^1$\\
     $H_{U(1)} - \delta$ & hypers charged\\
     & under $\widetilde{A}^1$ \\
     \multicolumn{2}{l}{+ KK towers of all fields} 
  \end{tabular}};
  
  \draw [-latex] (F_resolved.south) -> node[above, yshift=-3.5em, xshift=1.7em]{\begin{turn}{270}
 \begin{tabular}{c}compactify\\ on $S^1$\end{tabular}
 \end{turn}}(F_resolved_S1.north);
  
  \draw [-latex] (F_deformed.south) -> node[above, yshift=-4em, xshift=1.7em]{\begin{turn}{270}
 \begin{tabular}{c}compactify on\\$S^1$ with flux\end{tabular}
 \end{turn}}(F_deformed_S1.north);
  
  \kbbox[75, 160,north,M_resolved,(0, -16.5)] {\textbf{M-theory on $\hX$}};
    \node [] () at (-1.2, -16.3) {Massless sector:};
    \node [] () at (0, -17.1) {\begin{tabular}{ll}
     2 & gauge fields $A^a$ \\
     $H_{neutral}$ & neutral hypers \\
  \end{tabular}};
  
  \kbbox[75, 185,north,M_deformed,(10, -16.5)] {\textbf{M-theory on $\tX$}};
    \node [] () at (8.3, -16.3) {Massless sector:};
    \node [] () at (10.1, -17.1) {\begin{tabular}{ll}
     1 & gauge field $\widetilde{A}^0$ \\
     $H_{neutral} + \delta - 1$ & neutral hypers \\
  \end{tabular}};
 \draw [-latex] (F_resolved_S1.south) -> node[above, yshift=-4em, xshift=1.7em]{\begin{turn}{270}
 \begin{tabular}{c}integrate out \\ massive states \end{tabular}
 \end{turn}}(M_resolved.north);
 \draw [-latex] (F_deformed_S1.south) -> node[above, yshift=-4em, xshift=1.7em]{\begin{turn}{270}
 \begin{tabular}{c}integrate out \\ massive states \end{tabular}
 \end{turn}} (M_deformed.north);
 
 \draw [-latex] (F_resolved_S1.east) -> node[above, yshift=0em, xshift=0em]{\begin{turn}{0}
 \begin{tabular}{c}Conifold transition\end{tabular}
 \end{turn}}
 node[below]{\begin{tabular}{c}$\delta$ states with charge\\ $(q_0, q_1)$ become light \\ and Higgs gauge field\end{tabular}}
 (F_deformed_S1.west);
  \draw [-latex] (M_resolved.east) -> node[above, yshift=0em, xshift=0em]{\begin{turn}{0}
 \begin{tabular}{c}Conifold transition\end{tabular}
 \end{turn}}
 (M_deformed.west);
  
  \end{tikzpicture}
  \caption{A comprehensive summary of relations between the different theories and their spectra.}
  \label{fig:big_picture}
  \end{figure}

For completeness we have combined the figures of section \ref{sec:6d}, section \ref{sec:circle-flux},
section \ref{sec:eff_circle-flux}, and section \ref{sec:examples}
into figure \ref{fig:big_picture}, which summarizes the
relations between all the theories discussed in this paper.

Throughout the entire paper, we made extensive use of the information contained in the
Chern-Simons terms of the different five-dimensional theories that are generated
by integrating out charged massive matter fields. In order to perform validity checks
of the proposed F-theory models, we computed the matter spectra of the five-dimensional theories
corresponding to manifolds with section and tracked the Chern-Simons terms through the
transition to the manifolds that possess solely multi-sections.
Eventually, however, we were able to propose how to use the Chern-Simons terms to \emph{directly compute}
the matter spectra with respect to the massive $U(1)$s. The absence of an additional divisor
caused by the massiveness of the $U(1)$ appears to imply that one has somewhat less
control over F-theory models without section. Fortunately, though, it seems that one can use
anomaly cancelation to nevertheless compute the spectra without needing a conifold transition
to a model with section.

\subsection{Open questions and future directions of study}
Since the study of F-theory compactifications without sections is still a fairly unexplored topic,
there exists a plethora of ways to extend the results of \cite{Braun:2014oya, Morrison:2014era}
and this work. Given that the focus of these papers has been on the study of models with bi-sections,
it would certainly be desirable to have comparable control or at least access to a similar
number of example geometries with multi-sections of higher degree.

As discussed above, all of our example geometries in this paper are connected to Calabi-Yau manifold with
two sections by straightforward conifold transition. That this could be a general feature of such genus-one
fibrations is an enticing prospect. In particular, one can imagine that F-theory on genus-one fibrations
with 3- or 4-sections could be linked to F-theory models with multiple independent 1-sections by performing
not one, but several such conifold transitions, passing through models with, say, a 1-section and a 2-section
in intermediate steps.

Naturally, as just mentioned above, it would nevertheless be most convenient to access physical observables
of F-theory on manifolds without section directly -- that is without using additional related manifolds
such as the Jacobians or the manifolds obtained here by conifold transition. While we have demonstrated
for the explicit models studied above that this can be achieved, it still needs to be shown
that such an approach can be employed also for arbitrary gauge groups. Developing a general
framework for computing matter spectra under
the massive $U(1)$s or determining other physical observables would therefore
certainly be a promising direction of research.

Ultimately, to make contact with realistic F-theory models, one should extend the models studied here and in
\cite{Braun:2014oya, Morrison:2014era} to Calabi-Yau fourfolds without section. In principle, it is completely
straightforward to take the class of models studied in section \ref{sec:examples} and fiber the genus-one curves
presented there over a 3-complex-dimensional base instead. However, it would be interesting to examine
whether there exist additional features in F-theory compactifications to 4d and understand, for example,
whether the (non-)existence of certain Yukawa points has an impact on the states taking part in the conifold
transitions. In this context it may also prove useful to compute not only the chiral indices of the 4d
matter states, but rather their exact multiplicities using the formalism recently developed in \cite{Bies:2014sra}.

Lastly, we note that over the past years considerable effort (see for example \cite{Morrison:2012np, Morrison:2012js,
Grimm:2012yq, Martini:2014iza, Johnson:2014xpa}) has been made to systematically investigate
and classify six-dimensional supergravity models obtained from F-theory. In this approach, one usually considers
maximally Higgsed gauge groups and focuses on the remaining unbroken gauge group that a given base manifold
requires the overall fibration to have. At first sight, one might therefore expect not to detect the presence
of the sort of massive $U(1)$s treated in this paper. It would be interesting to see whether
there exists a way of nevertheless extracting such information.

\acknowledgments

We would like to thank Federico Bonetti, Volker Braun, James Gray,
Denis Klevers, David Morrison, and Washington Taylor for illuminating
discussions. I.G.-E. thanks N.~Hasegawa for kind encouragement and
constant support.

\appendix

\section{Geometric description of the matter multiplets in $\hX$}
\label{app:matter}

For the purposes of understanding the conifold transition, it was
sufficient to understand the $\rep{1}_4$ states in
table~\ref{table:resolved-CYs}. It is nevertheless interesting and
somewhat illuminating to describe the geometric origin of the rest of
the matter multiplets in the six-dimensional theory arising from
F-theory on $\hX$.

We start with the $\rep{1}_2$ multiplets. In fact, the relevant curves
have already been described in the $h^{1,1}=3$ cases explicitly
in~\cite{Morrison:2012ei} (under the names $\cT_n$, $0\leq n\leq
3$). We now review the discussion in that paper (using a slightly
different approach). Let us assume $f\neq 0$. We want to understand
under which conditions~\eqref{eq:resolved-quartic} factorizes into two
$\bP^1$s. This happens whenever the Calabi-Yau equation factorizes as
\begin{align}
  \label{eq:factorization-ansatz}
  \wt{\phi} = (w+B)(ws+C) = 0
\end{align}
for $B,C$ to be determined. For simplicity we restrict ourselves to
the case with $\deg(g)=0$, and set $g=1$. In this case, an easy
argument shows that a holomorphic redefinition of $w$ allows one to
set $\alpha=\beta=0$ in~\eqref{eq:P}. In what follows we will
implicitly perform such a redefinition.

Expanding~\eqref{eq:factorization-ansatz}, and comparing
with~\eqref{eq:resolved-quartic}, we immediately conclude that
\begin{align}
  \begin{split}
    BC & = y_1 Q'\\
    C+sB & = f y_2^2\, .
  \end{split}
\end{align}
By homogeneity and holomorphy, the most general form for $B$ is given
by
\begin{align}
  B = F y_1^2 s + G y_1 y_2
\end{align}
with $F,G$ polynomials in the $x_i$ variables of the appropriate
degree. (A term linear in $w$ is also possible, but this can be
reabsorbed in a redefinition of $w$.) Expanding the equations, and
comparing order by order, we arrive at the equations
\begin{align}
  b & = -F^2\\
  c & = -2FG\\
  d & = Ff - G^2\\
  e & = fG
\end{align}
which can be solved by
\begin{align}
  \begin{split}
    G & = \frac{e}{f}\\
    F & = \frac{1}{f^3}(df^2 + e^2)
  \end{split}
\end{align}
as long as
\begin{align}
  \begin{split}
    b & = -\frac{1}{f^6}(d^2f^4 + 2df^2e^2 + e^4)\\
    c & = -\frac{2}{f^4}(df^2e + e^3)\, .
  \end{split}
\end{align}
The $\rep{1}_2$ multiplets live at the points in the base where this
equation is satisfied. In order to count these points, we multiply the
whole equation by appropriate powers of $f$ (recall that $f\neq 0$ by
assumption), obtaining the equations
\begin{align}
  \label{eq:1-multiplet-loci}
  \begin{split}
    P_1 & \equiv bf^6 + d^2f^4 + 2df^2e^2 + e^4 = 0\\
    P_2 & \equiv cf^4 + 2df^2e + 2e^3 = 0\, .
  \end{split}
\end{align}
This set of equations has $(3\deg(e))(4\deg(e))=12\deg(e)^2$
solutions. Not all of these solutions correspond to ${\bf 1}_2$
states, though, some solutions come from $f=e=0$, which as discussed
in section~\ref{sec:examples} correspond to ${\bf 1}_4$ multiplets
instead. Each one of the solutions of $f=e=0$ contributes
$\deg_e(\Res_f(P_1, P_2))=16$ spurious solutions
to~\eqref{eq:1-multiplet-loci} (see \cite{Cvetic:2013nia}), so the
final count for ${\bf 1}_2$ multiplets is given by
\begin{align}
  H(\rep{1}_2) = 12 \deg(e)^2 - 16\deg(f)\cdot\deg(e)\, .
\end{align}
It is easy to check that this formula gives the right values for the
entries with $\deg(g)=0$ in table~\ref{table:resolved-CYs}.

Over the solutions of~\eqref{eq:1-multiplet-loci} with $f\neq 0$ in
the base, the elliptic fiber factorizes into the curves
\begin{align}
  \begin{split}
    c_B & = \{w + Fy_1^2s + Gy_1y_2 = 0\}\\
    c_C & = \{ws + fy_2^2 - F(sy_1)^2 - G(sy_1)y_2 = 0\}\, .
  \end{split}
\end{align}
The claim is that the hypermultiplets coming from wrapping $M2$ branes
on these curves have charge 2 under~\eqref{eq:DU(1)}. Notice first
that, since we are assuming $f\neq 0$, both sections are holomorphic,
and in particular $(c_B + c_C)\cdot \sigma_0 = (c_B + c_C)\cdot \sigma
= 1$, since the two components of the fiber, taken together, span the
class of the elliptic fiber. By the same token, the intersection is
transversal, so necessarily one of the intersections vanishes, and the
other is equal to 1. More explicitly, an easy calculation gives
\begin{align}
  c_B\cdot \sigma_0 = c_C\cdot \sigma=1\, ,\\
  c_B\cdot \sigma = c_C\cdot \sigma_0 = 0\, .
\end{align}
In addition, it is clear that $c_B\cdot [x_1] = c_C\cdot [x_1]=0$,
since the curves are localized over points in the base $\bP^2$, and
for the $g\neq 0$ case that we are considering there is no
intersection with the non-abelian divisor. All in all, we obtain that
$Q_{U(1)}=2$.

We now consider $H(\rep{2}_3)$. We claim that these hypers come
from the contracting spheres at $f=g=0$. As discussed above, over this
locus the $T^2$ fiber decomposes into three $\bP^1$ components. We
denote these components by $\cC_t$, $\cC_{y_1}$ and $\cC_\Xi$, and
claim that the ${\bf 2}_3$ hypers come from $\cC_{y_1}$ and $\cC_\Xi$
(the $M2$ states wrapping $\cC_t$ are rather associated with $W$
bosons of $SU(2)$).

Consider first $\cC_\Xi$. From the discussion above, we know that
$\cC_\Xi\cdot \sigma = 1$, $\cC_\Xi\cdot \sigma_0 = 0$ (since
$\sigma_0$ intersects the $\sigma$ rational component), $\cC_\Xi\cdot
[x_1]=0$ (by genericity) and $\cC_\Xi \cdot [t] = 1$. Plugging into
the charge formula, we conclude that $Q_{U(1)}=3$. In addition, the
$SU(2)$ Cartan is associated with $[t]$, so this is a charged state in
the fundamental, with charge one under the Cartan.

Similarly, for $\cC_{y_1}$ we have that $\cC_{y_1}\cdot \sigma_0 = 1$,
$\cC_{y_1}\cdot [x_1]=0$ and $\cC_{y_1}\cdot [t]=1$. The intersection
with $\sigma$ is again somewhat subtle, since $\sigma$ is rational,
wrapping the whole $\cC_{y_1}$. By the moving fiber argument,
$(\cC_{y_1} + \cC_\Xi + \cC_t)\cdot \sigma = 1$, and from $(\cC_\Xi +
\cC_t)\cdot \sigma=2$ we conclude that $Q_{U(1)} = -1$. Plugging these
values into the charge formula, we obtain $Q_{U(1)}=-3$. This state is
also charged under the $SU(2)$ Cartan with charge one. Taking the
conjugate state, we can complete the ${\bf 2}_3$ multiplet, as
advertised.

Let us now consider the $\rep{2}_1$ states. We consider
factorizations of the form
\begin{align}
  \label{eq:2_1-split-fiber}
  \wt{\phi} = t (b_0y_1s + b_1y_2)(b_2y_1^3 + b_3y_1^2y_2st +
  b_4y_1y_2^2t + b_5y_1ws + b_6y_2w)\, .
\end{align}
Here the $b_i$ are coefficients to be determined, and will depend on
the coefficients $b,c,\ldots$ of the Calabi-Yau equation. Such a
splitting exists whenever
\begin{align}
  \label{eq:2_1-locus}
  g(x_i) = I_1(x_i) = 0\, ,
\end{align}
with $I_1(x_i)=b^2f^3 + \ldots$ a certain polynomial of the $\bP^2$
coordinates $x_i$.\footnote{We computed~\eqref{eq:2_1-locus} by
  computing the elimination ideal associated to solving for the $b_i$
  variables in~\eqref{eq:2_1-split-fiber} in terms of the Calabi-Yau
  coefficients, using SAGE \cite{sage}.} This will hold at
\begin{align}
  \begin{split}
    \deg(g)\cdot \deg(I_1) & = \deg(g)\cdot(2\deg(b) + 3\deg(f)) \\
    & = -a^2 +
    3ab - 2b^2 + 12a - 9b + 45
  \end{split}
\end{align}
points in the base. Comparing with table~\ref{table:resolved-CYs} one
easily sees that this expression reproduces the $H(\rep{2}_1)$
multiplicities, so we expect that these hypermultiplets come from M2
branes wrapping these degenerations. Let us check this claim
explicitly.

Over a point satisfying~\eqref{eq:2_1-locus} we have that the fiber
degenerates, and in addition, generically $b_1\neq 0$
in~\eqref{eq:2_1-split-fiber}, since otherwise we would have three
polynomials intersecting over a point in $\bP^2$, which is
non-generic. We can thus locally redefine $y_2$ in such a way
that~\eqref{eq:2_1-split-fiber} becomes
\begin{align}
  \wt{\phi} = ty_2 (b_2s^2y_1^3 + b_3y_1^2y_2st + b_4y_1y_2^2t + b_5y_1ws
  + b_6y_2w)\, .
\end{align}
(This redefinition of $y_2$ is not necessary, but it simplifies the
presentation of the analysis.) Furthermore, comparing with the generic
form~(\ref{eq:restricted-quartic}) we can immediately identify
$b_4=e$, $b_6=f$, and similarly for the other coefficients. We see
that the fiber degenerates into three components: $\cC_t=\{t=0\}$,
$\cC_{y_2}=\{y_2=0\}$ and $\cC_{\Xi'}=\{b_2y_1^3 + \ldots\}$. Computing
the intersections amongst the components, and between the components
and the sections, is a completely straightforward exercise. The
resulting non-vanishing intersections are
\begin{align}
    \cC_t\cdot \cC_{y_2} = \cC_{y_2}\cdot\cC_{\Xi'} = \cC_{\Xi'}\cdot
    \cC_t = 1\\
    \cC_{\Xi'} \cdot P = \cC_t \cdot Q = 1\, .
\end{align}
Plugging into the charge formula~(\ref{eq:Q-Cs}), we obtain that the
M2 branes wrapped on $\cC_{y_2}$, $\cC_{\Xi'}$ form a doublet under
$SU(2)$ (since they are charged under the Cartan) with $U(1)$ charge
1, as expected from the counting above.

The last remaining set of states is $\rep{3}_0$. These have a somewhat
different origin. Notice that they are adjoints of the $SU(2)$ group,
this suggests that their origin comes from Wilson lines on the $SU(2)$
divisor, which we will call $\sG$. Recall that this divisor is given
by $\{g=0\}\subset\bP^2$, so its Euler character is, by adjunction:
\begin{align}
  \begin{split}
    \chi(\sG) & = \int_\sG c_1(T\sG) = \int_{\bP^2}[g]\wedge(3[x_1] -
    [g])\\
    & = \deg(g)(3 - \deg(g))
  \end{split}
\end{align}
or, equivalently, in terms of the genus $g_\sG$ of $\sG$
\begin{align}
  g_\sG = 1 - \frac{\deg(g)}{2}(3 - \deg(g))\, .
\end{align}
The $SU(2)$ Wilson lines on the (two) one-cycles associated with each
element of $g_\sG$, together with scalars coming from reduction of
$C_3$ on the same set of one-cycles (plus the contracting Cartan
divisor), one obtains exactly $g_\sG$ five-dimensional hypers in the adjoint
representation, which lift to $g_\sG$ six-dimensional hypers in F-theory. This
reproduces precisely the count displayed in
table~\ref{table:resolved-CYs}.

As an aside, let us highlight a small subtlety in checking six-dimensional anomaly
cancellation. If one naively plugs the matter content in
table~\ref{table:resolved-CYs} into the six-dimensional anomaly cancellation
conditions, one will see that the examples with $\rep{3}_0$ multiplets
do not satisfy gravitational anomaly cancellation. The explanation is
simple: deformations of $\sG$ can be described by complex structure
moduli variation of the total Calabi-Yau, i.e. elements of
$h^{2,1}(\tX)$, but they are also encoded in the values of the Wilson
lines over $\sG$. In particular, since the gauge group is $SU(2)$,
there is a single Casimir invariant, and each Wilson line degree of
freedom encodes one deformation modulus. We can see this a bit more
precisely: as emphasized in \cite{Beasley:2008dc}, for instance,
deformations of the $\sG$ locus are counted by sections the
anticanonical bundle $K_\sG$ of $\sG$, and using Serre duality
\begin{align}
  \dim H^0(K_\sG) = \dim H^1(\cO_\sG) = h^{0,1}(\sG)
\end{align}
which is precisely equal to $g_\sG$ for a connected Riemann surface,
such as $\sG$. All in all, in order to avoid overcounting one should
subtract $g_\sG$ neutral hypers from the contribution of
$h^{2,1}(\tX)$ to the gravitational anomaly, or alternatively count
the $\rep{3}_0$ multiplets with a multiplicity of 2, instead of 3.

\section{Non-existence of a section for $\tX$}
\label{sec:Oguiso}

In this appendix we would like to show that the deformed spaces $\tX$
considered in section~\ref{sec:examples} do not admit a section, but
rather a bi-section. I.e. there is no rational embedding of the base
$\bP^2$ into the total space such that the fiber is generically
intersected at a single point. The best that we can do is finding
divisors of the total space that project down to the base, but
generically intersect the fiber twice, i.e. a bi-section. The basic
idea was described in \cite{Oguiso,Morrison:1996na}.

In order to prove this, we need to identify the fiber curve
first. This is easy, it is simply given by $\cT=[x_1]^2\cap \tX$. This
is intuitively easy to understand: the fiber is obtained by taking the
preimage of a point (with class $[x_1]^2$) in the base $\bP^2$.

Now we need to prove that there is no section $\sS$. In all of our
examples, the K\"ahler cone of the Calabi-Yau $\tX$ can be generated
by the restrictions of the toric divisors $[x_1]$, $[y_1]$, and in the
cases with $h^{1,1}(\tX)=3$, also $[w]$. We thus parametrize
\begin{align}
  \label{eq:S-parameterization}
  \sS = a[x_1] + b[y_1] + c[w]
\end{align}
with coefficients (a priori not necessarily integral) to be
determined. The generic intersection between the $T^2$ fiber and the
section is given by
\begin{align}
  \cT\cdot \sS = 2b + 4c\, .
\end{align}
Showing that this can never be equal to one would follow if
$b, c \in\bZ$. This is indeed the case, as we now show. Consider first
the case with $h^{1,1}(\tX)=3$, since it is somewhat simpler. Over a
locus in the base given by
\begin{align}
  g(x_i) = I_2(x_i) = 0
\end{align}
with\footnote{As in appendix~\ref{app:matter} this is obtained using
  SAGE~\cite{sage}.}
\begin{align}
  \begin{split}
    I_2(x_i) & = f^{4} b^{2} -  \beta f^{3} b c + \alpha f^{3} c^{2} +
    \beta^{2} f^{2} b d - 2 \alpha f^{3} b d -  \alpha \beta f^{2} c d
    + \alpha^{2} f^{2} d^{2} -  \beta^{3} f b e\\
    &\phantom{=} + 3 \alpha \beta f^{2} b e + \alpha \beta^{2} f c e -
    2 \alpha^{2} f^{2} c e -  \alpha^{2} \beta f d e + \alpha^{3} f
    e^{2} + \beta^{4} b a - 4 \alpha \beta^{2} f b a \\
    & \phantom{=} + 2 \alpha^{2} f^{2} b a -  \alpha \beta^{3} c a + 3
    \alpha^{2} \beta f c a + \alpha^{2} \beta^{2} d a - 2 \alpha^{3} f
    d a -  \alpha^{3} \beta e a + \alpha^{4} a^{2}\, ,
  \end{split}
\end{align}
the Calabi-Yau equation~\eqref{eq:quartic} factorizes into three
factors
\begin{align}
  \phi = t(b_0 y_1 + b_1 y_2)(b_2y_1^3 + b_3y_1^2y_2t +
  b_4y_1y_2^2t + b_5y_2^3t + b_6y_1w + b_7y_2w)\, .
\end{align}
The important part for our analysis is that this defines three
holomorphic curves in the Calabi-Yau: $\cC_t=\{t=0\}$, $\cC_y=\{b_0
y_1 + b_1 y_2=0\}$ and $\cC_\Xi$ for the other component. (The
notation is intended to be reminiscent of that used in
appendix~\ref{app:matter}. Indeed, the matter we just found is
precisely the $\rep{2}_1$ and $\rep{2}_3$ multiplets on the resolved
side taken together, since after the Higgsing of the $U(1)$ they
cannot be separated anymore.) Computing the intersection numbers with
the generators of the K\"ahler cone chosen
in~\eqref{eq:S-parameterization} is an easy exercise, we get
\begin{align}
  \begin{split}
    \cC_t\cdot [y_1] & = 1\, \\
    \cC_y\cdot [w] & = 1
  \end{split}
\end{align}
with all other intersections vanishing. Since the intersection between
a divisor and a curve in a smooth space has to be integral, by
intersecting $\sS$ with these curves we conclude that $b,c\in\bZ$, and
thus $\cT\cdot\sS\in 2\bZ$. I.e. there is no section, but rather a
bi-section.

This argument fails for the cases with $h^{1,1}(\tX)=2$, since $g=0$
has no solutions. From the previous discussion it is nevertheless
clear what to do, though: the $\rep{1}_2$ states on the resolved side
$\hX$ that we described in appendix~\ref{app:matter} will survive the
conifold transition, and appear on the deformed side $\tX$ as loci on
the $\bP^2$ base where the fiber degenerates as
\begin{align}
  \phi = (w + B)(w + D)\, .
\end{align}
Computing the intersection numbers one gets
\begin{align}
  \begin{split}
    \cC_B\cdot [y_1] & = \cC_C\cdot [y_1] = 1\\
    \cC_B\cdot [x_1] & = \cC_C\cdot [x_1] = 0
  \end{split}
\end{align}
and since a putative section $\sS=a[x_1]+b[y_1]$ has intersection
$\sS\cdot\cT=2b$ with the fiber $\cT$, this shows that indeed we have
no section, but rather a bi-section.

\bibliographystyle{JHEP}
\bibliography{refs}

\end{document}